 \documentclass[twocolumn]{aastex631}

\newcommand{\GAL}{GaLNet}
\def\lsim{\mathrel{\rlap{\lower3.7pt\hbox{\hskip0.5pt$\sim$}}
    \raise0.5pt\hbox{$<$}}} 
\def\gsim{\mathrel{\rlap{\lower3.7pt\hbox{\hskip0.5pt$\sim$}}
    \raise0.5pt\hbox{$>$}}}

\graphicspath{{./}{figures/}}

\begin{document}

\title{Galaxy Light profile neural Networks (GaLNets). II. Bulge-Disc decomposition in optical space-based observations}

\author{Chen Qiu}
\affiliation{School of Physics and Astronomy, Sun Yat-sen University, Zhuhai Campus, 2 Daxue Road, Xiangzhou District, Zhuhai, PR China}

\author[0000-0003-0911-8884]{Nicola R. Napolitano}\footnote{{\rm napolitano@mail.sysu.edu.cn}\\{\rm liruiww@gmail.com}}
\affiliation{School of Physics and Astronomy, Sun Yat-sen University, Zhuhai Campus, 2 Daxue Road, Xiangzhou District, Zhuhai, PR China}
\affiliation{CSST Science Center for Guangdong-Hong Kong-Macau Great Bay Area, Zhuhai, 519082, PR China}

\author{Rui Li}
\affiliation{School of Astronomy and Space Science, University of Chinese Academy of Sciences, Beijing 100049, PR China}
\affiliation{National Astronomical Observatories, Chinese Academy of Sciences, 20A Datun Road, Chaoyang District, Beijing 100012, PR China}

\author{Yuedong Fang}
\affiliation{University Observatory, Faculty of Physics, Ludwig-Maximilians Universität, Scheinerstr. 1, 81679 Munich, Germany}

\author{Crescenzo Tortora}
\affiliation{
INAF—Osservatorio Astronomico di Capodimonte, Salita Moiariello 16, I-80131—Napoli, Italy}

\author{Shiyin Shen}
\affiliation{Key Laboratory for Research in Galaxies and Cosmology, Shanghai Astronomical Observatory, CAS,
Shanghai, 200030, PR China}
\affiliation{ Key Lab for Astrophysics, Shanghai, 200034, Shanghai, PR China}

\author[0000-0001-6947-5846]{Luis C. Ho}
\affiliation{Kavli Institute for Astronomy and Astrophysics, Peking University, Beijing 100871, China}
\affiliation{Department of Astronomy, School of Physics, Peking University, Beijing 100871, PR China}


\author{Weipeng Lin}
\affiliation{School of Physics and Astronomy, Sun Yat-sen University, Zhuhai Campus, 2 Daxue Road, Xiangzhou District, Zhuhai, PR China}
\affiliation{CSST Science Center for Guangdong-Hong Kong-Macau Great Bay Area, Zhuhai, 519082, PR China}

\author{Leyao Wei}
\affiliation{School of Physics and Astronomy, Sun Yat-sen University, Zhuhai Campus, 2 Daxue Road, Xiangzhou District, Zhuhai, PR China}

\author{Ran Li}
\affiliation{National Astronomical Observatories, Chinese Academy of Sciences, 20A Datun Road, Chaoyang District, Beijing 100012, PR China}

\author{Zuhui Fan}
\affiliation{South-Western Institute for Astronomy Research, Yunnan University, Kunming 650500, Yunnan, PR China}

\author{Yang Wang}
\affiliation{CSST Science Center for Guangdong-Hong Kong-Macau Great Bay Area, Zhuhai, 519082, PR China}
\affiliation{Peng Cheng Laboratory, No.2, Xingke 1st Street, Shenzhen, 518000, PR China}

\author{Guoliang Li}
\affiliation{Purple Mountain Observatory, Chinese Academy of Sciences, 2 West Beijing Road, Nanjing 210008, China}

\author{Hu Zhan}
\affiliation{Key Laboratory of Space Astronomy and Technology, National Astronomical Observatories, Chinese Academy of Sciences, 20A Datun Road, Beijing 100101, PR China}
\affiliation{Kavli Institute for Astronomy and Astrophysics, Peking University, Beijing 100871, China}

\author{Dezi Liu}
\affiliation{South-Western Institute for Astronomy Research, Yunnan University, Kunming 650500, Yunnan, PR China}


\begin{abstract}
Bulge-disk (B-D) decomposition is an effective diagnostic to characterize the galaxy morphology and understand its evolution across time. So far, high-quality data have allowed detailed B-D decomposition to redshift below 0.5, with limited excursions over small volumes at higher redshifts. Next-generation large sky space surveys in optical, e.g. from the China Space Station Telescope (CSST), and near-infrared, e.g. from the space EUCLID mission, will produce a gigantic leap in these studies as they will provide deep, high-quality photometric images over more than 15000 deg2 of the sky, including billions of galaxies. Here, we extend the use of the Galaxy Light profile neural Network (GaLNet) to predict 2-S\'ersic model parameters, specifically from CSST data. We simulate point-spread function (PSF) convolved galaxies, with realistic B-D parameter distributions, on CSST mock observations to train the new GaLNet and predict the structural parameters (e.g. magnitude, effective radius, Sersic index, axis ratio, etc.) of both bulge and disk components. We find that the GaLNet can achieve very good accuracy for most of the B-D parameters down to an $r$-band magnitude of 23.5 and redshift $\sim$1. The best accuracy is obtained for magnitudes, implying accurate bulge-to-total (B/T) estimates. To further forecast the CSST performances, we also discuss the results of the 1-S\'ersic GaLNet and show that CSST half-depth data will allow us to derive accurate 1-component models up to $r\sim$24 and redshift z$\sim$1.7.
\end{abstract}



\section{Introduction} \label{sec:intro}

Despite their variegate morphology, most of the physical 
processes behind galaxy evolution
can be 
understood by the detailed study of their two main stellar components: bulges and disks. According to the most credited 
galaxy formation scenario,
disks are formed by cooled gas from dark matter halos (\citep{1978MNRAS.183..341W}; \citep{1980MNRAS.193..189F}) and bulges are generally generated from the merging of two disks  (\citep{1967MNRAS.136..101L}; \citep{1972ApJ...178..623T}; \citep{1977egsp.conf..401T}; \citep{1991ApJ...370L..65B}; \citep{1988ApJ...331..699B}; \citep{2006MNRAS.369..625N}), or unstable gas-rich disks ({\citep{2004ARA&A..42..603K}}, \citep{2016ASSL..418..355B}). 
Accurately characterizing bulges and disks in galaxies is 
a difficult but necessary step to reveal their formation and evolutionary history (e.g., \citep{2005ApJ...620..564C}; \citep{2014ApJ...788...11Lang14}; \citep{Gao_2017}).

In particular, parametric fitting, describing the surface brightness profile of the different galaxy components with their structural parameters (e.g. the magnitude, the effective radius, etc.), has long been proven to be a powerful tool in galaxy analysis (\citep{1959HDP....53..275D}; \citep{1968adga.book.....S}; \citep{1977ApJ...218..333K}).
Traditionally, 
multi-component galaxies are represented by an exponential disk (\citep{1970ApJ...160..811F}) with a \citet{1948AnAp...11..247D} bulge (see e.g. \citep{1995MNRAS.275..874A_exp-devauc}), although it has been found that  \citet{1968adga.book.....S} profiles with $n$-index, i.e. the central slope in the projected light, larger than the de Vaucouleurs's $n=4$ can better reproduce the bulge components combined to exponential disks (\citep{2019ApJS..244...34Gao19}).

A more general approach adopts a 
a S\'ersic
profile to model both components. Here, bulges and disks can be distinguished by the S\'ersic index, 
being generally $n\lsim2$ for disks, and $n\gsim2$ for the bulge/spheroids (\citep{2003MNRAS.343..978Shen+2003}, \citep{2005PASA...22..118G}, \citep{2008AJ....136..773F}). 
In this case, one can
reproduce the surface brightness distribution of galaxies 
with a more realistic combination of bulge-disc components (\citep{2004bdmh.confE..83M}). 
Ideally, the 2-S\'ersic models 
works well for bright/large galaxies, but it is harder for faint/small systems due to their low signal-to-noise ratio (SNR) even in a relatively nearby universe
(\citep{2006MNRAS.371....2Allen2006}, \citep{2022MNRAS.516..942C}). For high redshift galaxies, this becomes even harder due to the limited number of pixels with sufficiently high SNR to include in their modeling, hence requiring the use of space observations to best perform this kind of analysis (\citep{2014MNRAS.444.1660B}).

Cosmic epochs at z$\sim$1 and beyond 
are crucial for galaxy morphology evolution as the star-formation rate
density reaches its peak in the {universe's} history (\citep{2014MNRAS.444.1660B};  {\citep{2014ARA&A..52..415M}}), and the categories of bulges, disks and spheroids {experience} epochal transformations (\citep{2021ApJ...913..125C}). For this reason, we are motivated to push surface photometry studies of galaxies to improve our understanding of their evolution processes (see e.g. \citep{2005ApJ...620..564C}; \citep{2022arXiv221001110F}). {This is particularly} important to fully test the predictions from cosmological hydro-dynamical simulations, which are providing unprecedented details on the internal structure of individual galaxies at different epochs (e.g., 
\citep{2015MNRAS.454.1886S}; \citep{2017MNRAS.467.2879B,2017MNRAS.467.1033B}; \citep{2018ApJ...853..194D}; \citep{2019MNRAS.483.4140R}; \citep{2020ApJ...895..139D}). 
So far, detailed modeling of the multi-component structure of galaxies have been limited to low redshift (\citep{2011ApJS..196...11Simard_B/D_SDSS}; \citep{2013ApJ...766...47H}; \citep{2020ApJS..247...20Gao2020}), with only few space-based programs dedicated to high-redshift samples, over small areas and statistics (\citep{2010ApJ...721..193P}, \citep{2014MNRAS.444.1660B})
However, with 
the upcoming large sky surveys from space (e.g. Chinese Space Station Telescope -- CSST, \citep{2019ApJ...883..203G}; {Euclid} mission, \citep{2011arXiv1110.3193L}; Roman Space Telescope -- Roman, \citep{2020arXiv200805624M}), we have the chance to move to high quality data over large volumes up to high redshifts, while deeper ground based programs (e.g. from the Vera Rubin LSST, \citep{2019ApJ...873..111I}) will also provide exquisite data for extremely faint and/or diffuse systems in the local universe. This unprecedented data collection will improve our understanding of galaxy morphological transformation
up to z$>$1, {in samples with over a billion galaxies.}

Unfortunately, galaxy surface photometry represents a bottle-neck
of this learning process because of the time demand of traditional galaxy fitting methods.
{Tools based on 2D galaxy surface brightness distribution measurements,
like GALFIT (\citep{2002AJ....124..266P}), Gim2d (\citep{1998ASPC..145..108S}), and 2DPHOT (\citep{2008PASP..120..681L}), PROFIT (\citep{2017MNRAS.466.1513R}), have been extensively used to measure structural parameter in galaxies, either as single- or multi-component systems (\citep{2013MNRAS.433.1344M}, \citep{2020ApJS..247...20Gao2020}, \citep{2023arXiv230308627X}).}
However, these traditional codes are either too slow or need too much manual intervention to be suitable for
billion galaxy samples.
Thus, even though most of these codes can reach a fairly high accuracy ``if'' initial conditions are correctly given (\citep{2011MNRAS.414.1625Y}), more automatic methods with similar or even higher accuracy are demanded.

Among all options, Machine Learning (ML) tools are becoming a game changer in the approach to big dataset analysis and interpretation. 
In the last decade ML tools have been regularly applied in a variety of research areas, including {astronomy}. They can easily perform tasks like classification or regression with unprecedented speed and accuracy and they have been used in the analyses of gravitational waves (\citep{2015arXiv150205037C};
\citep{2013PhRvD..88f2003B}), the photometric classification of {supernovae} (\citep{2016ApJS..225...31L}), the search for strong gravitational lenses (\citep{2017MNRAS.472.1129P,2019MNRAS.484.3879P,2019MNRAS.482..807P}; \citep{2019yCat..22430017J}; \citep{2020A&A...644A.163C}, \citep{2020ApJ...899...30L, 2021ApJ...923...16L}), star/galaxy classification (\citep{2021A&A...645A..87B}), unsupervised feature-learning for galaxy SEDs (\citep{2017A&A...603A..60F}) and galaxy morphology classification (\citep{2010arXiv1005.0390G}; 
\citep{2004MNRAS.348.1038B};
\citep{2010MNRAS.406..342B}). 
Convolutional Neural Networks (CNNs) are a particular 
class of ML tools that are designed to derive features from arrays of data by convolution, and, as such, are optimal in images processing.
For instance, CNNs {have} been used in galaxy classification (\citep{2015MNRAS.450.1441D}) and surface brightness distribution analysis (\citep{2018MNRAS.475..894T}; \citep{2020SPIE11452E..23U}; \citep{2022ApJ...929..152L}). 

In this context, we have started developing the GAlaxy Light profile convolutional neural Networks (GaLNets, \citep{2022ApJ...929..152L}, Li+22 hereafter) to perform, for the first time, single S\'ersic profile fitting of galaxies from ground-based data with the accurate treatment of the point-spread function (PSF).
Compared to traditional tools, the GaLNets can reach similar accuracies
with a computational speed more than three orders of magnitude faster.
As a first application of GaLNets to ground-based galaxy surface photometry, we have used a sample of galaxies from the Kilo Degree Survey (KiDS: \citep{2015A&A...582A..62D,2017A&A...604A.134D}; \citep{2019A&A...625A...2K}) and shown that CNNs can effectively and accurately perform single S\'ersic analyses of very large samples of galaxy light profiles from the ground, {similar to the ones} that will be collected by VR/LSST. 
The upcoming all-sky space observations
motivate us to extend the use the GaLNets 
to perform detailed 2-component analysis of galaxies over a wide redshift range. So far, the only similar attempt, we are aware of, from \citet{2021MNRAS.506.3313G} is limited to the derivation of the bulge-to-total (B/T) luminosity of nearby galaxies. 

In this paper, we adopt a similar scheme of the first GaLNets (Li+22) to implement a PSF-convolved, 2-S\'ersic profile fitting of galaxies in space observations. In particular, we use the case of the CSST, which is explicitly optimized in the optical wavelengths, and in particular, we will concentrate on the {single-epoch} $r$-band data sample, as the reference dataset for this first test (see more details on \S\ref{sec:ccst_simu}). As in Li+22, we simulate 2-dimensional mock galaxies, but here we use two component (bulge and disk) systems using parameters from the CosmoDC2 catalog (\citep{2019ApJS..245...26K})\footnote{This is a large synthetic galaxy catalog made for VR/LSST simulated datasets, which covers 440$\rm deg^{2}$ of sky area to a redshift $z=3$.}, except the S\'ersic index, that in CosmoDC2 is assumed 1 and 4 for the disk and bulge respectively. Instead, we took a more realistic distribution of the S\'ersic indexes of the two components from \citet{2016MNRAS.460.3458K}. 
We assumed a typical {\it PSF} from the CSST to convolve 2D bulge-disk S\'ersic models that we add to randomly selected "background cutout" from CSST mock observations, and finally obtain realistic galaxy mock observations. 
Finally, we have trained the GaLNets on 1-component galaxies to evaluate the applicability of these networks  to a more general variety of real observations with CSST.


This work is organized as follows. In Sect.2, we describe how to build the training and testing sample and describe the CNN architectures.
In Sect.3, we test our CNNs on simulated data. In Sect.4, we discuss the results of the GaLNets and in Sect.5 we draw our main conclusions.

\section{Methods and Data} \label{sec:style}

CNNs have been inspired by the research of the visual cortex of biological brains (\citep{hubel_wiesel_1962}; \citep{fukushima_1980}; \citep{lecun_bengio_hinton_2015}). 
In contrast to conventional Neural Networks that consist of fully-connected layers and tend to disregard the underlying data structure, CNNs make use of architectures that preserve the pertinent information encoded within the data's inherent structure.
In particular, they use so-called Convolution layers, that have the ability to carryover ``feature'' information (e.g. a pattern or a color in an image) by reducing the size of the elements containing relevant data.
This allows the CNNs to remarkably save computational power on those applications, like image processing, dealing with large arrays of data.
Hence, CNNs are very efficient at making predictions of certain target features (i.e. a pattern, a property, one or more parameters) over specific objects in high-resolution astronomical images and spectra, and can be used either in classification (e.g. galaxy morphology, \citep{2020MNRAS.491.1554W}; strong gravitational lensing search, \citep{2019MNRAS.482..313L,2020ApJ...899...30L,2021ApJ...923...16L}) or regression problems (e.g. for spectroscopic feature recognition and redshifts, \citep{2016A&C....16...34H}, \citep{2022RAA....22f5014Z}; galaxy fitting, \citep{2018MNRAS.475..894T}, \citep{2022ApJ...929..152L}).
To make accurate predictions, though, it is indispensable that 1) the data used for the CNN training (training set)  
realistically reproduce the ones used to make predictions (predictive set) and 2) training sets do cover the full target parameter space.

Ideally, these conditions can be satisfied by collecting a large sample of real systems labelled by the ``true'' features one wants to predict on a new dataset with unknown targets (i.e. the astrophysical parameters). This is practically out of the reach as 1) we cannot know the true parameters of astrophysical objects but only determine them via other parametric or non-parametric tools, which are naturally prone to biases; 2) even if we assume that we can estimate bias-free ``targets'' with standard analysis tools, very often this process is time-consuming and the final training set would be too small and noisy, to prevent systematics.  
The use of simulated data is a very common solution to obviate these shortcomings, provided that the process of producing mock data takes all the observational and physical parameters correctly into account. 
\subsection{A GaLNet for Bulge-Disk decomposition: {GaLNet-BD}} \label{sec:galnet}
As introduced in Sect.4, we want to apply the GaLNets to a classic regression task where the inputs are images of bulge-disk galaxies and the corresponding PSFs, and the outputs are the parameters of the best 2-S\'ersic profiles describing the 2D galaxy light distribution (see below). Since this new GaLNet is specialized for B-D decomposition, we have dubbed it GaLNet-BD. The S\'ersic (1968) profile is defined as: 

\begin{equation}
     I(x,y)= I_{\rm e}{\rm exp}\left \{ -b_{n}\left[
     \left(\frac{\sqrt{q(x-x_{\rm 0})^{2}+(y-y_{\rm 0})^{2}/q}}
     {R_{\rm e}}\right)
     ^\frac{1}{n} -1\right] 
     \right \} 
     \label{eq:sersic}
\end{equation}
where 
$R_{\rm e}$ is the effective radius, $I_e$ is the surface brightness at the effective radius, $q$ the axis ratio, $n$ is the S\'ersic index, while ($x_{\rm 0}$, $y_{\rm 0}$) are the coordinates of the center. We also define the position angle, $PA$, which represents the angle between the minor galaxy axis and the North to East direction on the sky. 
In Eq. \ref{eq:sersic}, 
for $b_{n}$ we use the expression provided by \citet{1999A&A...352..447C}:
\begin{equation}
b_{n} \approx\left\{
\begin{array}{lr}
             2n-1/3+4/(405n) , & n\ge0.36  \\
             0.01945-0.8902n+10.95n^{2}  & n\textless0.36.
             \end{array}\right.
\end{equation}

Eq. \ref{eq:sersic} is used to describe both the Bulge and Disk components, and let us to define the total surface brightness of the B-D system as
\begin{equation}
     I_{\rm tot}(x,y)= I_{\rm Bulge}(x,y)+I_{\rm Disk}(x,y). 
     \label{eq:Itot}
\end{equation}

The total (apparent) magnitude, $mag$, is defined by
\begin{equation}\label{mag}
    \centering
    mag = -2.5 \log (F_{tot})+zpt
\end{equation}
where $zpt$ is the zero point of the adopted filter.
$F_{tot}$ is the total flux of the galaxies, definded as

\begin{equation}\label{mag}
    F_{tot}=2\pi \sum_{i=1}^2 R_{{\rm e}, i}^2 I_{{\rm e}, i} e^{b_{n_{i}}} n_{i} b_{n_{ i}}^{-2n_{i}} \Gamma (2n_{i}) q_{i},
\end{equation}

where $i=1, 2$ is for bulge and disk and $\Gamma$ indicates the standard $\Gamma-$function. 

\begin{figure}
\centering
\vspace{-0.4cm}
\includegraphics[width=0.43\textwidth ]{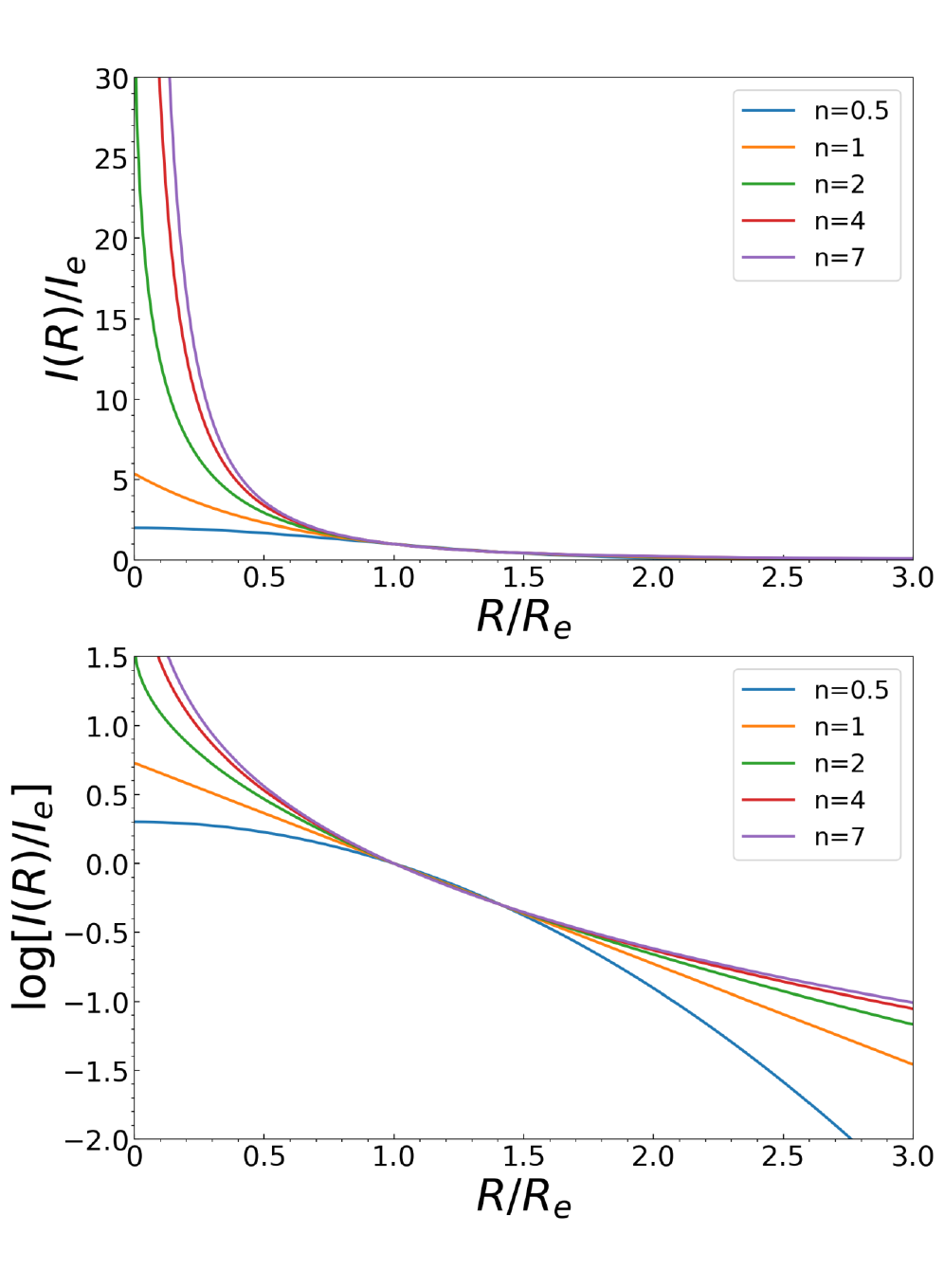}
\caption{1D normalized S\'ersic surface brightness profiles in linear (top) and ``log'' scale (bottom) as a function of the radius in units of $R_{\rm e}$, for different $n$-indices ($n=0.5,1,2,4,5,7$, from blue to purple).
}
\label{fig:sersic_fig}
\end{figure}

As the S\'ersic index is known to be a proxy of the galaxy morphology (e.g. \citep{2007ASSP....3..481T}), we assume $n<$ 2 for the thin disk component, while for the bulge we generally assume $n>$2, but also include $n<2$ (pseudo-bulges,
e.g. \citep{2008AJ....136..773F}, \citep{2016MNRAS.460.3458K}). 
In Fig. \ref{fig:sersic_fig}, we show the dependence of the 1D profiles as a function of the $n$-index. 
In particular we see that the S\'ersic profiles with large $n$-index ($n>2$, i.e. typical of the dominant ``bulges'') are clearly distinct from the ones with $n<2$ in the central regions, although they become blended for larger and larger $n$ value. This is also true for the large $n$-index in the outer regions (i.e. $R>R_{\rm e}$) where all profiles look almost indistinguishable, while it is easier to separate the models with lower $n$-index ($n<2$). 
This gives a perspective of the difficulty, in B-D decomposition, one 
needs to overcome when accurately predicting the overall parameters. As we will see, this is important for our deep learning methods, as they ``look'' at the central regions, to learn about bulges, and the outer regions, to learn about disks.

\begin{figure*}[t]
\centering
\plotone{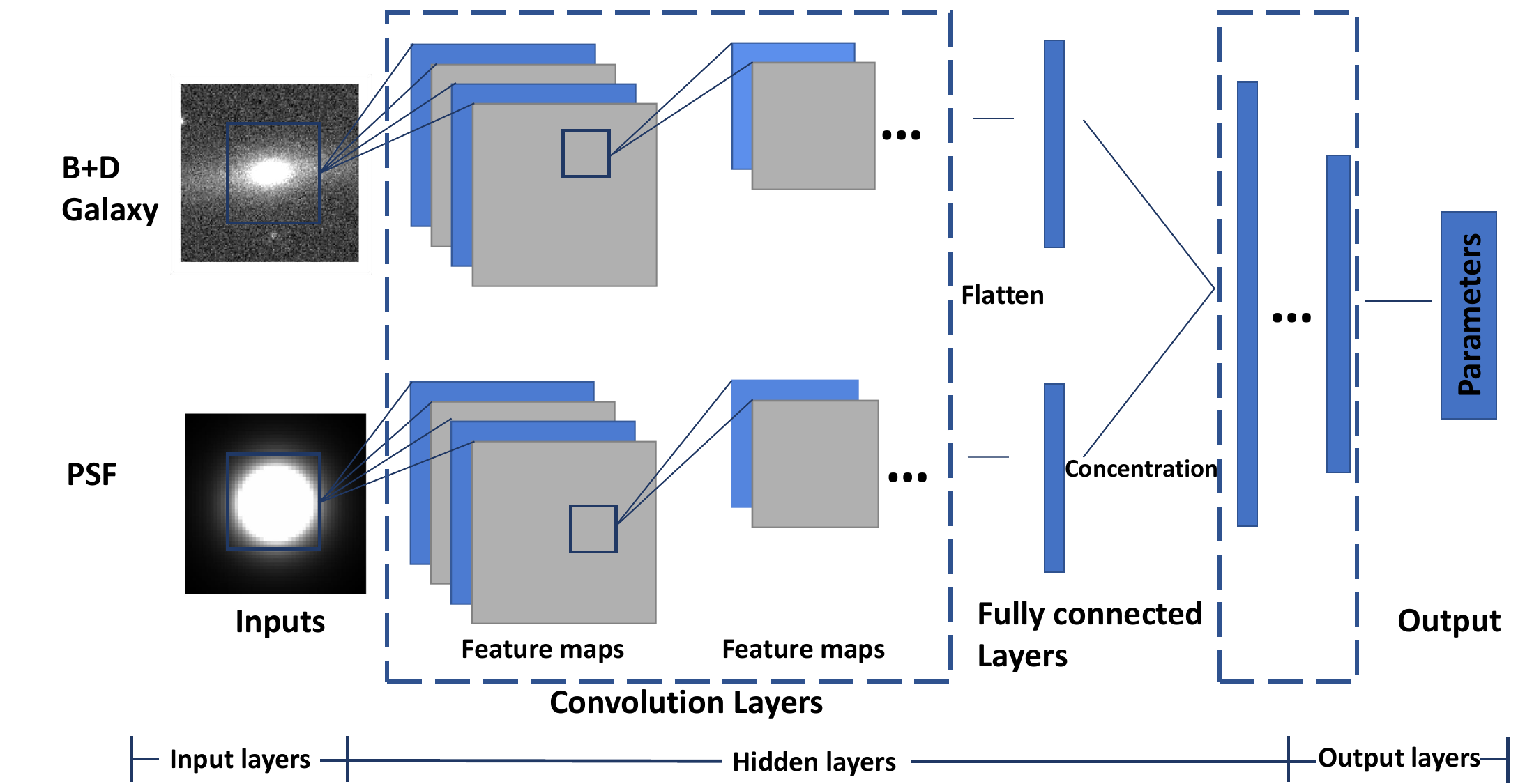}
\caption{The structure of the GaLNet-BD used in this work. The networks are fed by both galaxy images and the corresponding “local” simulated PSFs, and the outputs are the 10 parameters of the two S{\'e}rsic profiles. 6 layers are used for the galaxy branch and 4 layers for the PSF branch. After the Concatenation layer, we add 3 fully connected layers to extract further features of the combined galaxies and PSFs.
\label{fig:general}}
\end{figure*}

The total number of free parameters of the fitting process for a B-D decomposition is 14, i.e. the 7 parameters $x_{\rm cen}$, $y_{\rm cen}$, $mag$,  $R_{\rm e}$,  $q$,  $n$,  $PA$ for each of the two components. As we keep the center of the two components fixed to the central pixel of the image cutout, we are left with {\it 10 parameters} to be constrained by the GaLNet-BD.

In order to reproduce real observations we need to account for the local background, $BG$, of typical instrumental observations, which are dependent on a series of observational factors, including the cosmic light background, 
the detector noise, and the {\it PSF}, which also depends on the optical design and observational conditions.  
All these factors are combined to produce the mock galaxies, using 
the equation:

\begin{equation}
     I_{\rm Obs}(x,y)= I_{\rm tot}(x,y)  \otimes PSF +I_{BG}(x,y)
\label{eq:Iobs}     
\end{equation}
where $I_{\rm tot}(x,y)$ is the total galaxy surface brightness distribution from Eq. \ref{eq:Itot}, and $I_{BG}(x,y)$ is the value of local background, while $\otimes$ denotes convolution. More details on this procedure will be given in \S\ref{sec:sim_gal}.

Thanks to the high quality expected from the space images of CSST 
degeneracies among the S\'ersic parameters in Eq.s \ref{eq:sersic} and \ref{eq:Itot} are expected to be less severe than ground-based observations (see, e.g. \citep{2001MNRAS.321..269T}). In particular, space-based instruments tend to have 
relatively stable {\it PSF}, determined only by instrumental effects such as diffraction due to obscuration, optical aberrations, and polishing errors, as well as thermal variations in the telescope structure (see e.g. the breathing effect of Hubble Space Telescope). It is therefore theoretically possible to model the {\it PSF} based on the instrument optics (\citep{2011SPIE.8127E..0JK}), but there are other observational parameters difficult to account. To overcome this, 
we assume the {\it PSF} following a Gaussian distribution with FWHM centered at 0.15$''$ and a scatter, $\sigma_{\rm FWHM}=0.015''$, which conservatively accounts for the variation of the {\it PSF} across the FOV of the CSST camera 
(see \S\ref{sec:floats}, for more details).


\subsection{The CNN Architecture} \label{sec:style}


CNNs are characterised by a weight sharing network structure, which reduces the complexity of the network model and the number of weights. This is very advantageous in two-dimensional image processing, avoiding the complex feature extraction and data reconstruction process of the traditional recognition algorithm. 
GaLNets are built based on ``supervised'' learning: the input data are simulated galaxy images,
labeled with corresponding S\'ersic parameters (labels).

In this work, we apply an adapted VGGNet (\citep{simonyan_zisserman_2014}). Its structure is a canonical CNN, consisting of three parts: the input layer, the hidden layer, and the output layer (see Fig. \ref{fig:general}). The core of the CNN is the hidden layer, which includes 1) the convolutional layer made of multiple convolution kernels composed of weights and biases, which can be used in extracting features from the input data 2) the pooling layer, and 3) the fully connected layer. 
The “feature maps” derived by the convolutional layer are passed to the pooling layer, which performs some information filtering to reduce the features' sizes. The pooling layer contains a preset pooling function used to replace the feature map in a given region with a single point (flattening). Finally, the fully connected layer, located at the end of the hidden layer, performs a nonlinear combination of the extracted features. In particular, it combines the low-level learned features into high-level features and passes them to the output layer. In Fig. \ref{fig:general}, we show how in the GaLNet-BD, we apply this structure to two different channels that are meant to extract the features from two inputs in parallel, the galaxy image and the {\it PSF} image. In order to combine the information selected in the two channels the GaLNet is equipped with a concentration layer, right before the fully connected layer, where the parameters are predicted on the basis of the weights imposed by the {\it PSF} branch.

\subsection{Data: Mock CSST galaxy cutouts} \label{sec:floats}
According to the architecture shown in the previous paragraph, our training set consists of 1) 2D simulated bulge-disk galaxy images and 2) the corresponding 2D PSFs. To build this data set, we add 2-dimensional, {\it PSF}-convolved, simulated bulge-disk galaxies to randomly selected $r$-band cutouts from CSST simulations, representing the galaxy “background”. In this section, we give a short description of the CSST simulated observations we have used, and the process followed to produce the mock galaxy cutouts.

\begin{figure*}[t]
\centering
\plotone{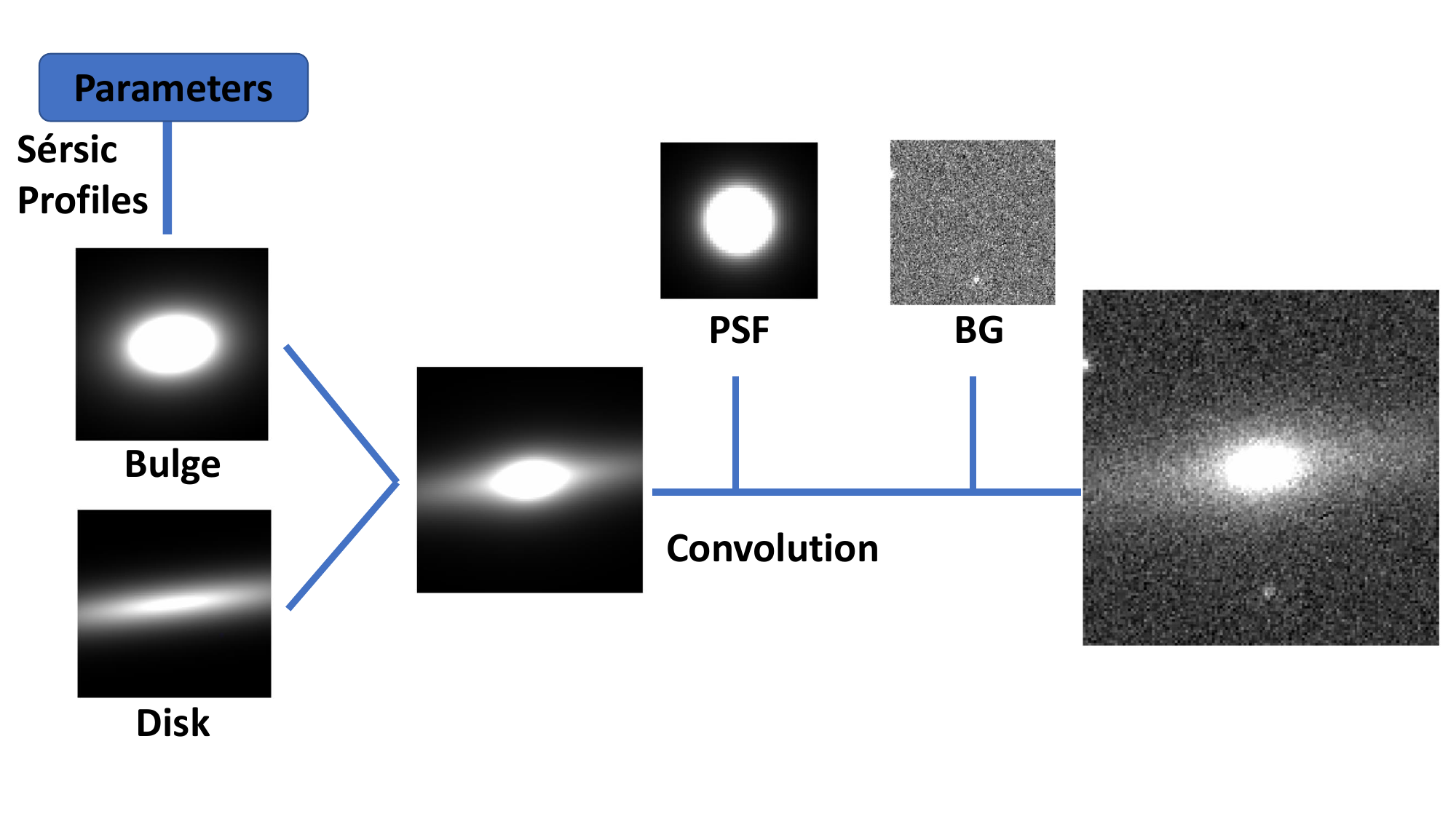}
\caption{Structural and positional parameters are used in fitting S\'ersic model for both bulges and disks. To make realistic simulations, we combined 2-S\'ersic model convolved with {\it PSF} and CSST background value (see \S\ref{sec:sim_gal}).}
\label{fig:b+d_sim}
\end{figure*}

\subsubsection{CSST simulations}
\label{sec:ccst_simu}
In this work, we will use the China Space Station Telescope (CSST) as a prototype facility for large sky space observations in optical bands. The CSST is a 2-m telescope capable of observations in 7 bands from the UV to optical/NIR (i.e. NUV$ugrizy$) using a wide field camera covering an effective area of $\sim1.1$ deg$^2$ with a pixel scale of $0.074\rm ~arcsec/pxl$. It will carry out a wide survey over 17\,500 deg$^2$ and a deep survey over 400 deg$^2$ with $r\sim26$ and $r\sim27.2$ limiting mag, respectively.
The superb image quality (FWHM$<0.15''$) and photometric depth ($r<26$) will be ideal for performing
dark matter tomography and constraining the dark energy equation of state ({\citep{2023A&A...669A.128L}}). However, CSST will also provide unprecedented multi-band data to study the structure of galaxies and AGN and their evolution, across {cosmic} time. CSST is expected to be launched by mid-2025 and since the camera has not yet finalized, there are currently no real images to be used as a templates for this work. 

However, in preparation of CSST operations and to empirically assess the performance of design and hardware, the CSST Science Ground Segment has implemented a suite of software  
based on GALSIM \citep{2015A&C....10..121R} framework to produce simulated images {(Fang Y., private communication)}. Despite these simulations are not optimized for science, they provide a realistic dataset to produce mock observations, suitable for science tests.

The code is made publicly available\footnote{ https://csst-tb.bao.ac.cn/code/csst\_sim/csst-simulation} and can be used to generate pixel-level CSST exposures to different fidelity level. 
The workflow of this simulation can be summarized by the following stages:\\

1) Truth catalogues, survey strategies, {\it PSF} samples, and field distortion model are prepared separately from the imaging simulation. In particular, the super-sampled {\it PSF} stamps are calculated over the focal plane, and in four colours within each CSST bandpass via ray-tracing in CODE-V\footnote{ https://www.synopsys.com/optical-solutions/codev.html}. \\

2) {\it PSF} stamps are further interpolated in GALSIM to get a spatially-varying, quasi-chromatic {\it PSF} model. To simulate a single exposure, in each of the exposures, each galaxy is modeled as the sum of an exponential disk and a De Vaucouleurs bulge, and each star is modeled by a simple Dirac Delta function. Photon flux is assigned to each object according to its magnitude, SED, and the corresponding filter. Locations of objects are given by projecting their celestial coordinates on to image coordinates via WCS and field distortion model.\\

3)  In each filter, the surface brightness profiles of objects are convolved with the {\it PSF} model at their locations. Rendering is handled by the “photon-shooting” option in GALSIM. Stamps from all sub-bandpasses are stacked, and various detector effects are modeled and added to get the final image.\\

The simulated ``raw'' data produced by the CSST simulation group have been processed by the current version of the CSST pipeline. A detailed description of the pipeline will be provided in a dedicated paper (Fang Y., private communication). Generally speaking, this pipeline performs a chip-to-chip standard data reduction, including bias and overscan subtraction, flat-fielding, astrometric and photometric calibration, cosmic ray subtraction.

For this paper, we make use of single epoch observations, consisting in a 150$s$ exposure for which we have
measured a limiting magnitude for extended sources of $r\sim {25.3}$\footnote{This has been obtained by comparing the number of observed galaxies with respect to the input catalog, as a function of the $r$-mag and determine the magnitude where 50\% of the input galaxies are lost.}. 
We have chosen the single-epoch because the full depth for the CSST wide survey will be available for the whole wide-survey area only at the end of the 10-year mission of the satellite, while 
the single-epoch will be the reference dataset available for the first years of telescope operations. Hence, it is worth demonstrating that galaxy structural parameters of galaxies and B-D decomposition are feasible also with half of the nominal depth of the survey. We will address the full depth of the wide survey and the deep field areas forecasts in forthcoming analyses.

\subsubsection{Simulating realistic B-D galaxy images}
\label{sec:sim_gal}
In this section, we describe in more detail the process to produce realistic multi-component galaxies made of bulges and disks. As seen in the previous paragraph, the CSST simulated images contain simple galaxies made of an exponential disk and a De Vaucouleur profile, which are not representative of real systems (see \ref{sec:galnet}). Rather, we intend to use a more physical distribution of this parameter in our analysis (see below, and also \S\ref{sec:intro}). Overall, the simulated images described in the previous section represent the most realistic template of what the CSST observations will look like and provide the necessary realism to our test in terms of background light, image distortions and companion systems, we might eventually encounter in our analysis. 

The overall simulation procedure is sketched in Fig. \ref{fig:b+d_sim}, and described in details here below. \\

\begin{figure*}[t]
\centering
\plotone{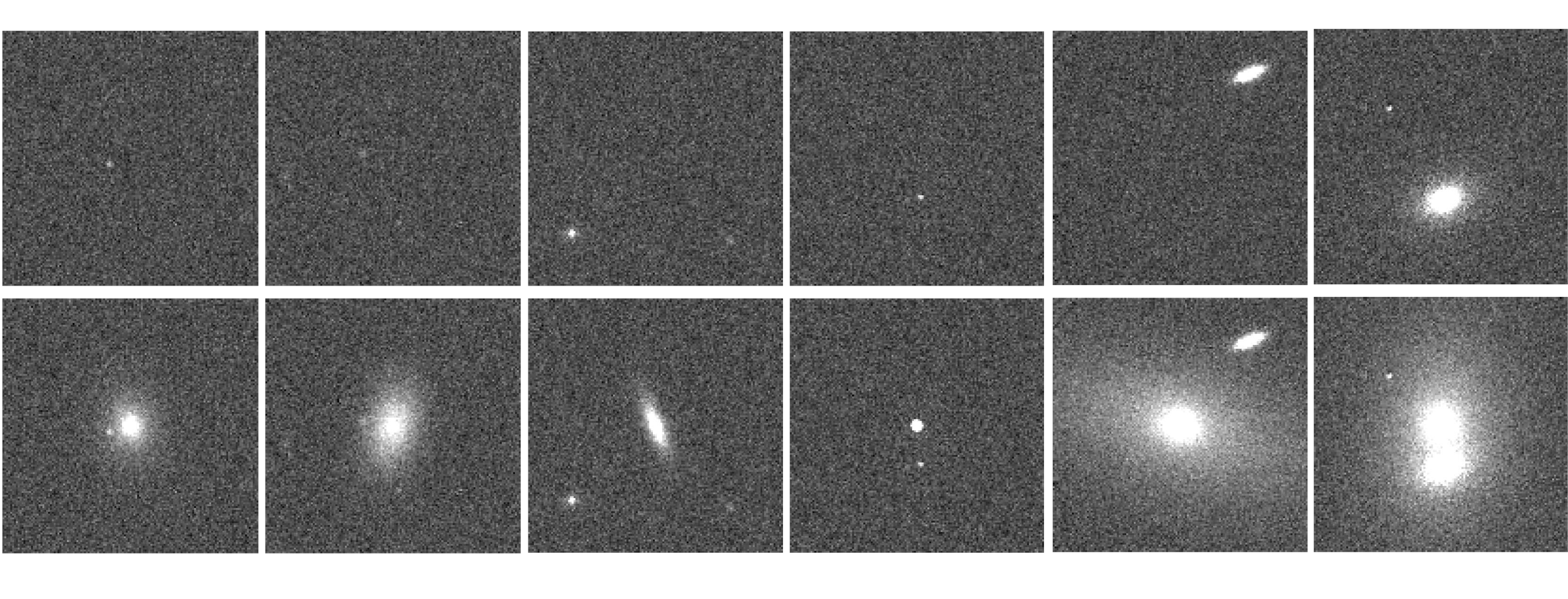}
\caption{A set of reduced $r$-band CSST simulated images (first row) and the same images with simulated galaxies according to the procedure illustrated in \S\ref{sec:sim_gal} (second row). }
\label{fig:sim_sample}
\end{figure*}

1)  {\it Background images}. We randomly select small cutouts of size 135$\times$135 pixels corresponding to about 10$\times$10 arcsec$^2$ from $r$-band CCDs  of the CSST simulated images. 
To provide a realistic environment for the mock galaxies, we have allowed these ``background'' cutouts to contain other simulated sources, like stars and galaxies, with the further addition of cosmic rays. 
The only attention we use, at this stage, is to remove those cutouts with too bright source (galaxies or stars) in the central region. We finally produce $\sim$ 1200 galaxy ``background'' cutouts. 
In Fig. \ref{fig:sim_sample} (top) we show a small sample of these images, where we can clearly distinguish the structure of the background noise and also the presence of simulated sources, from faint compact to bright extended sources. We further increase this sample via ``augmentation'', by applying a 90, 180, 270 deg rotation and flipping, to finally collect 9600 images. 
\\

2)  {\it Galaxy magnitudes}. As anticipated in \S\ref{sec:intro}, to simulate S\'ersic profiles, we use a realistic catalog of mock observations from cosmological N-body simulations as the one produced for the VR/LSST (CosmoDC2; \citep{2019ApJS..245...26K}), which contains galaxy parameters up to redshift $z=3$. In CosmoDC2, each 2-component galaxy is characterized by a multitude of galaxy properties including stellar mass, morphology, spectral energy distributions, broadband filter magnitudes, host halo information, and weak lensing shear. For our work, we are interested on so-called ``structural parameters'' for the two components, including luminosity, effective radius, axis ratio and position angle, for different photometric bands, namely $ugrizy$, although for this first test we will use only the $r$-band\footnote{We need to remark that the CosmoDC2 mock catalogs are tailored to VR/LSST observations and the $r$-band filter, as well as the other filters, might slightly differ from the ones that will be used by CSST. We expect though that the differences are insignificant to try to correct these for possible zero points. We also remark that this color term does not impact our analysis as the train and test samples will make use of the same calibration.}, while we leave the multi-band analysis making use of the other bands for future work (see also \ref{sec:perf_linear} for a discussion). 
Regarding magnitudes, the CosmoDC2 provides the total luminosities of the bulge and disk components of each galaxy, and the conversion formula:
\begin{equation}
     mag_r = -2.5* \log (L_r)-2.5* \log(1+z)+\mu(z) 
\end{equation}
where $L_r$ is the $r$-band luminosity of either component, while $z$ is the redshift and $\mu(z)$ is the distance modulus of the galaxy.

Even though CosmoDC2 contains parameters of $\sim2$ billions of galaxies in the 440 deg$^2$ simulated area, most of them are too faint to be observed by CSST, which is $\sim$ 1.5 mag fainter than VR/LSST, if considering the wide survey. This difference is 
worsened as we are using only half of the total depth of the final CSST wide survey program. We will come to the issue of completeness in \S\ref{sec:SNR_cut}.\\

3)  {\it Other S\'ersic profile parameters}. 
To make predictions of the 10 parameters of the two-component S\'ersic model discussed in \S\ref{sec:galnet}, we need to create a training sample realistically reproducing the expected distribution of intrinsic galaxy parameters. In Table \ref{table:parameters} we give a summary of the parameters adopted in the simulation of the training sample as derived from the CosmoDC2 catalog. Here we remind that, for our training sample, we have adopted a $n$-index distribution wider than the one of the original CosmoDC2 catalog, and assumed the one from an observational sample from \citet{2016MNRAS.460.3458K}. In particular, we see that the $n$-index of the bulges, $n_{\rm bulge}$, span from {\bf $n_{\rm bulge}\sim1$ }(pseudo-bulges) to $n_{\rm bulge}=8$, while the $n$-index of the disks, $n_{\rm disk}$, are canonically smaller than 2. We also remark that the CosmoDC2 mock catalogs do not have any conditions on the relative size of bulges and disks, including also cases where the effective radius of the bulges is larger than the ones of the disks (embedded disks). Although in principle we could adopt some empirical relation to bind bulge and disk sizes (see e.g. \citep{2014ApJ...787...69D}), we decide here to leave more freedom to these particular prior distributions. Note that these latter are not the final ``observed'' parameter distributions, as the imposition of the SNR cut for accurate surface photometry will produce a further selection of the mock sample parameters (see Sect. \ref{sec:SNR_cut}). 

The parameters, as in Table \ref{table:parameters}, are randomly sampled and used in Eqs. \ref{eq:sersic} and \ref{eq:Itot} to model the simulated galaxies that will be used as training and test samples. The only parameters that we decided to keep fixed because of minor physical meaning are the galaxy centers ($x_{\rm cen}$, $y_{\rm cen}$). These are assumed to be (0,0), i.e. the center of the ``background'' image. 
Here below we illustrate the  steps to produce the mock images of these simulated galaxies.

\begin{deluxetable}{llll}
\tablenum{1}
\tablecaption{\\Parameters for Simulating the Training Samples }
\tablehead
{Parameter & Range &Unit&Distribution }
\startdata
Mag       & 17-24                &mag     & Given by DC2   \\
$R_{\rm e,bulge}$; $R_{\rm e,disk}$   & 0.2-6                  &arcseconds  & Given by DC2  \\
q         & 0.02-1                 &\nodata     & Given by DC2  \\
pa        & 0.00-180               &degrees     & Given by DC2 \\
$x_{\rm cen}$   & 0                      &pix     & set  \\
$y_{\rm cen}$   & 0                      &pix     & set \\
$n_{\rm bulge}$ & 0.3-8                  &\nodata     & F \\
          &                        &            & (n=30,d=5) \\
$n_{\rm disk}$  & 0.5-2                  &\nodata     & Normal\\
          &                   &            &($\mu=1$,
          $\sigma=0.5$) \\
\enddata
\tablecomments{Range and distribution of parameter values used to simulate the galaxies. $\mu$ and $\sigma$ are the mean value and standard deviation of a normal distribution. n is the degrees of freedom in the numerator and d is the degrees of freedom in the denominator.}
\label{table:parameters}
\end{deluxetable}

4) {\it PSF}. Accurately accounting for the {\it PSF} is a crucial pre-requisite for unbiased structural parameter estimates (\citep{2001MNRAS.321..269T}), even for space observations (\citep{2020MNRAS.496.5017G}). 
To do that, for our mock galaxies we make use of self-made ``Gaussian PSF images'' (see Fig. \ref{fig:general}), although these might be slightly different of the ones resulting from the ray tracing for the CSST simulations.  
For sake of generality, to produce a Gaussian profile we assume a circularly symmetric, Moffat-like profile, but with parameters reproducing a Normal distribution. In particular we adopt the following equation  (\citep{2001MNRAS.328..977T}):

\begin{equation}
{\it PSF}(r)=\frac{\beta-1}{a^{2}\pi}[1+(\frac{r}{a})^{2}]^{-\beta}
\end{equation}

where $r$ is the distance from the center of the ``{\it PSF} image'' in pixels, and where we choose $\beta$=100 (\citep{2001MNRAS.328..977T}), corresponding to a Gaussian distribution. The choice of the Gaussian profile
is motivated by ray tracing tests on the CSST optics showing that the {\it PSF} can be approximately Gaussian (Fang Y., private communication). According to these tests, the FWHM in $r$-band is conservatively close to 0.1$''$ with $\sim$10\% fluctuation. However, due to charge diffusion on the CCD, the ``observed'' FWHM can become significantly larger and asymmetric. To check that, we directly measure a dozen of ``simulated stars'' in CSST images, obtaining a mean FWHM$=0.150''\pm0.005''$, with no signs of significant ellipticity and a minimal presence of tails deviating from a Gaussian in the outermost {\it PSF} profiles.
Finally, to add more realism to the {\it PSF} effect on the mock galaxy images, we assume some CSST chip-by-chip variation, by taking the local Gaussian {\it PSF} having the FWHM drawn from a Gaussian distribution with mean FWHM$=0.150''$ and variance $\sigma({\rm FWHM})=0.015''$. The size of these ``PSF images'' is 51$\times$51 pixels, or 3.8$\times$3.8 arcsec$^2$, which is wide enough to fully sample the FWHM of the PSF. Note that, at this stage, the details of the {\it PSF} model are secondary, as these do not impact the accuracy of the predictions: as we have discussed elsewhere (Li+22), to maximize the accuracy the GaLNets need to learn the local PSF, regardless the kind of model adopted. This latter can be changed for the training and the direct modeling of real galaxies when the real {\it PSF} of the instrument will be measured on real images.\\

5) {\it {\it PSF} convolution and final images.} This is the step where we convert a 2D, 2-component galaxy model into a realistic mock observation of the galaxy sample. As anticipated in \ref{sec:galnet}, we first convolve the 2D model in Eq. \ref{eq:Itot} with the {\it PSF} and then, after having added Poisson noise to the convolved profile, we add the resulting 135$\times$135 pixel (i.e. 10$''\times10''$) image to the 
background cutout. We remark here that this provides an area large enough to sample most of the light profile of galaxies with $R_{\rm e}$ $\sim$  $1''$ (corresponding to $\sim$ 14 pixels). For larger galaxies, though, the fraction of the total light enclosed in the cutout can be smaller than the total one. As discussed in Li+22, this does not constitute a problem as the CNN is sensitive to the light gradients in the brighter regions, while it looses sensitivity in the low surface brightness, low SNR regions. Here, given the range of $R_{\rm e}$ adopted in this paper, mostly $<2''$, with a small fraction of systems having $R_{\rm e}>2''$ at low-$z$ (see next \S\ref{sec:SNR_cut}), we still expect to sample generally an area enclosing $2-3 R_{\rm e}$s, which is enough to clearly separate different profiles. 
Overall, the final cutout size was decided as a compromise between area sampling and computational speed, as this latter is a non-linear function of the cutout size, especially for data reading/writing.\\

This latter step produces a typical real-looking galaxy, as shown in Fig. \ref{fig:sim_sample} (bottom), for which we can measure the signal-to-noise ratio to verify if this is large enough to perform accurate surface brightness analysis.  
In particular, this is measured over a central area, covered by the  
effective radius ($R_{\rm e}\times R_{\rm e}$).

In this paper, 
we use ${\rm SNR}=40$, 
which is slightly smaller than previous GaLNets' experiments based on ground-based observations in Li+22, but large enough to obtain a reasonable accuracy for the structure parameters (see also \citep{simonyan_zisserman_2014}, \citep{2018MNRAS.480.1057R}). We will eventually evaluate the impact of the lower SNRs in the parameter predictions in future analyses, with the aim of pushing the completeness limit toward the smaller fluxes (see also \S\ref{sec:SNR_cut}). 

In Fig. \ref{fig:sim_sample} (bottom), we can also see the variety of ``blending situations'' we have included in our training sample, realistically accounting of the presence of compact stars and extended galaxies, often overlapping with the simulated systems, whose centers are, by construction, coincident with the cutout center.

%



\begin{figure}

\hspace{-0.7cm}
\includegraphics[width=9.5cm]{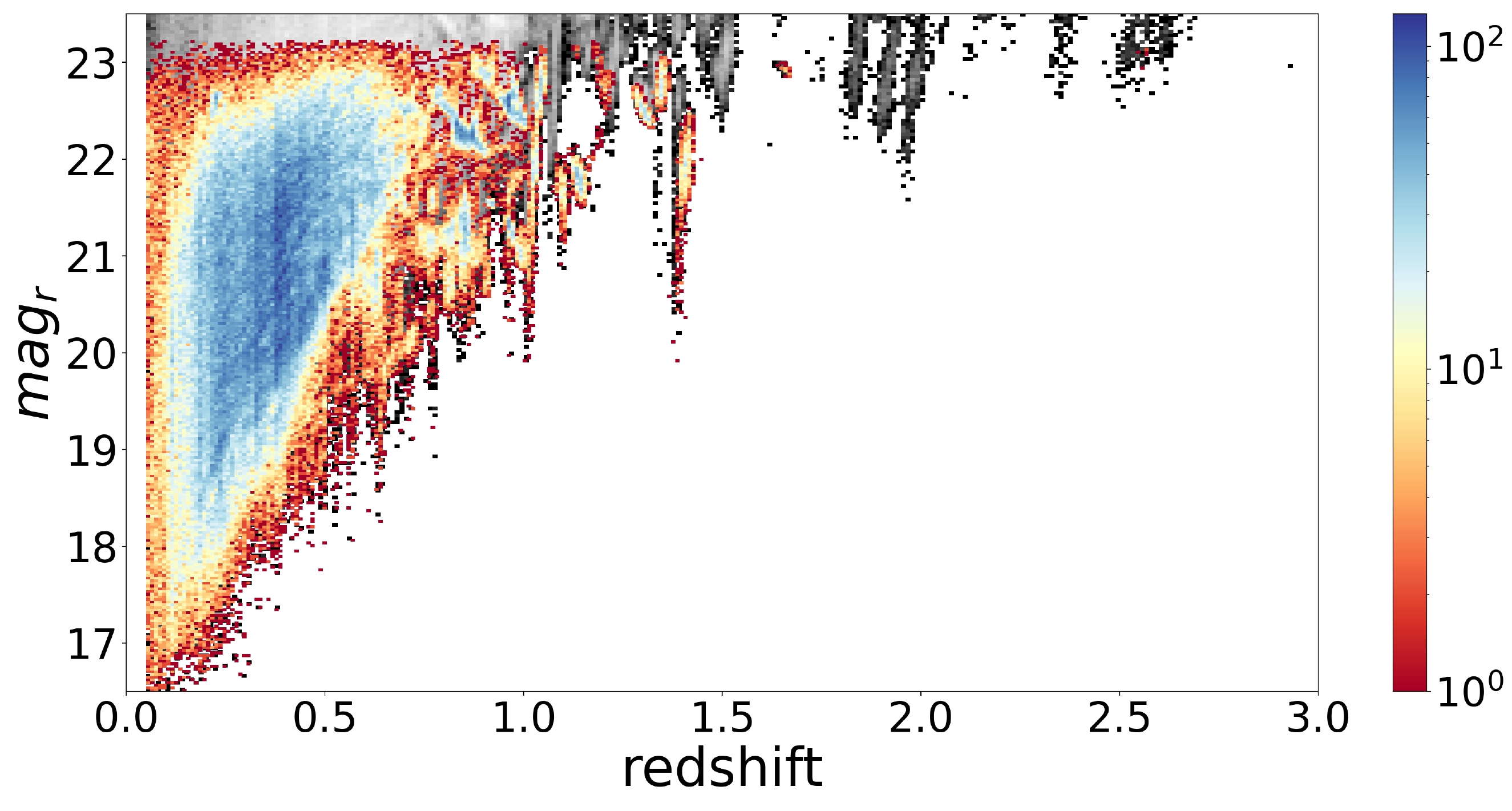}
\caption{Redshift versus total magnitude before (gray points) and after (color + bar) the SNR cut. We can see that the SNR cut produces a drop in the galaxies accessible to our analysis at $r>23$. Also, galaxies at high redshift ($z>1.5$) have too low SNR ($<40$) to also qualify for the 2-component fitting. \label{fig:z-mag}}
\end{figure}

Following steps 1–5 above, we simulate 250\,000 mock galaxies with redshift up to 1.5 (according to Table \ref{table:parameters}). Every simulated galaxy, to be selected as part of this mock galaxy sample, has to pass the ${\rm SNR}=40$ criterion (see also \S\ref{sec:SNR_cut} here below for a discussion on the selection effects). We finally split this sample into the training data, made of 200\,000 mock galaxies, of which 40\,000 galaxies are used for the validation, and the test sample, made of the residual 50\,000 simulated galaxies. 

\begin{figure*}
\centering
\vspace{-0.cm}
\includegraphics[width=1.0\textwidth ]{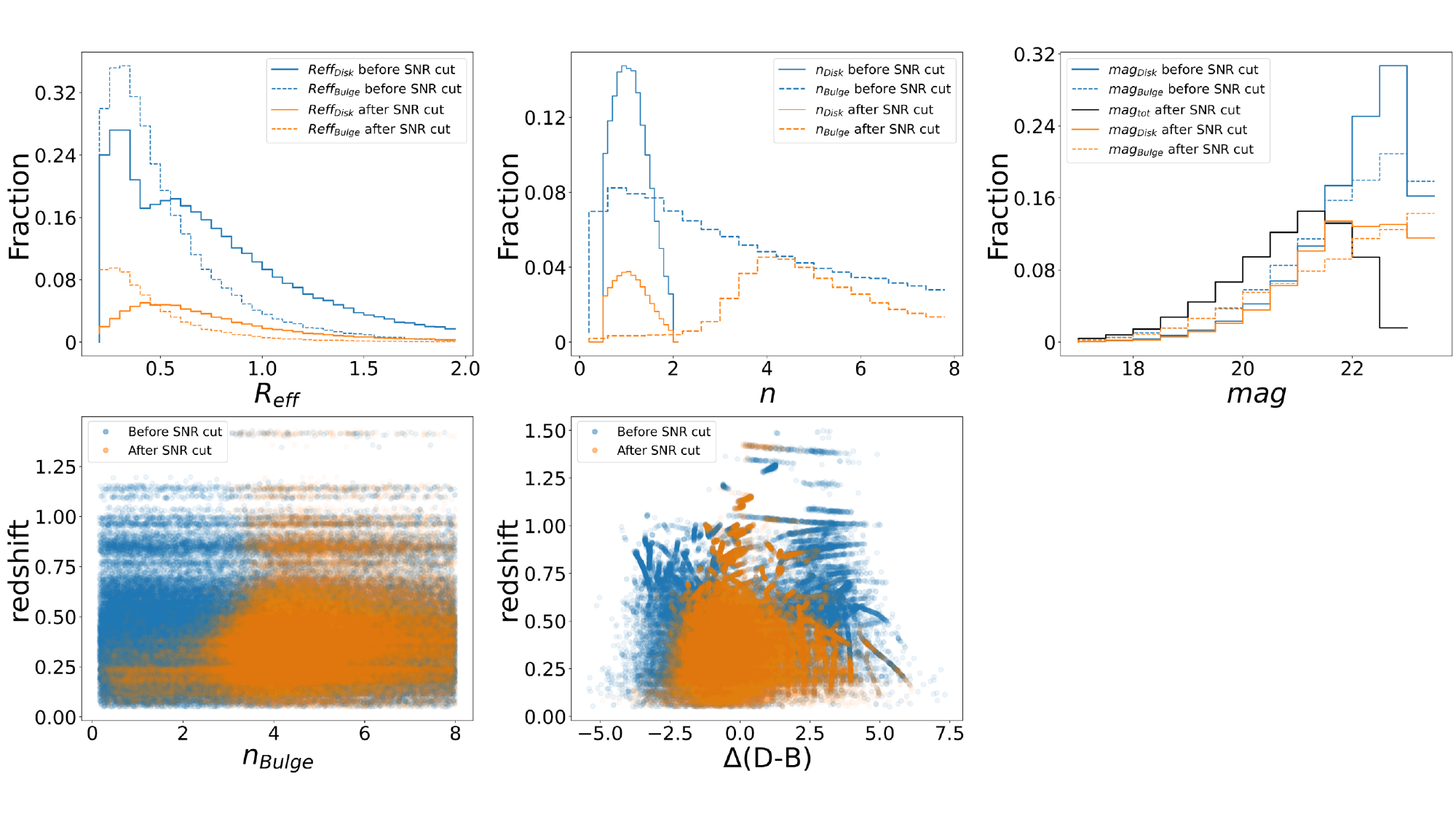}
\caption{Distribution of the mock galaxy parameters before (blue) and after (orange) the SNR cut. Top row: disk and bulge distributions of $R_{\rm e}$, $n$, $mag$, from left to right show that SNR = 40 criterion suppress especially the small sized disks, the small $n_{\rm bulge}$ and the low-luminosity disks. All other parameters maintain the same distribution behavior. The black histogram on the magnitude plot represents the distribution of the total magnitudes of the B+D systems, showing the confusion limit at $r\sim23$ (see text). Bottom row: the redshift distribution of the $n_{\rm bulge}$ and the Disk/Bulge ratio as measured by the $\Delta$(D-B) parameter (see text). Here we see (left panel) that only a small part of bulges with $n_{\rm bulge}$ survive the SNR cut. We also see (right panel) that disks-dominated ($\Delta$(D-B)$<-2.5$), suffer a strong selection at all redshifts, while bulge-dominated-disks ($\Delta$(D-B)$>2.5$) pass the SNR=40 criterion, while they become looser at $z>0.5$.  
\label{fig:n_distr}}
\end{figure*}

\subsubsection{SNR selection and final parameter distributions}
\label{sec:SNR_cut}
In Fig. \ref{fig:z-mag} we show the distribution of the galaxy magnitude as a function of redshift of the galaxy obeying to the $\rm SNR=40$ criterion. 
From the plot, we see that the number of galaxies fainter than $r=23$ suddenly drops at all redshifts, despite the input distribution reaches $r=24$ (see Table \ref{table:parameters}), showing that this is a fair approximation of the ``completeness'' limit of 
our B-D decomposition with one epoch CSST observations, for SNR=40. Below this limit, we do not attempt, at this stage, to perform any surface brightness analysis. 
In the same figure we also see that for $z>1$, galaxies with SNR$\leq$40 are sparsely distributed, and almost disappear at $z>1.5$.
Hence, 
we keep only galaxies at $z<1.5$ for the B-D modeling.   
Besides the magnitude, the SNR cut has an impact also on the distribution of all the other parameters. This is shown in Fig. \ref{fig:n_distr}, where we plot the $n$-index, $r$-mag and $R_{\rm e}$ for bulges and disks. Here, starting from  
magnitudes, we notice that the distribution of disks becomes flatter at $r>21.5$, meaning that faint disks are difficult to model and are suppressed in our final statistics. We also see that the distribution of the $R_{\rm e}$ of the disks is changed, with a stronger suppression of the $R_{\rm e}>0.3$ and in particular of the peak at $R_{\rm e}\sim0.5$. Overall, the stronger effect we can see is on the $n_{\rm bulge}$ distribution. In particular, bulges with $n_{\rm bulge}<3$ (including pseudo-bulges) tend to be selected off because their more diffuse profile produces lower SNRs. 
This ``selection effect'' is possibly not a major issue for our analysis as both the training and test samples will follow the same parameter distributions (but see Appendix). However, it rings a bell about the application to real galaxies, where the ``selection effect'' from SNR limitation can 
produce an incompleteness of intermediate/high redshift pseudo-bulges (see top-left panel of Fig. \ref{fig:n_distr}). The predominance of high $n_{\rm bulge}$ has been found in previous analysis (e.g. \citep{2006MNRAS.371....2Allen2006}), which confirms that this 
is a realistic feature incorporated in our training sample. 
Finally, following the strong effect on $n_{\rm bulge}$ we also show the impact of the SNR on the selection of the Disk-to-Bulge ratio, measured via the difference of the disk and bulge magnitudes, $\Delta$(D-B)$=mag_{\rm disk}-mag_{\rm bulge}$ in $r$-band. Here we see that disk-dominated systems disks-dominated ($\Delta$(D-B)$<-2.5$), suffer a strong selection at all redshifts, except at $z\lsim0.1$ while bulge-dominated-disks ($\Delta$(D-B)$>2.5$) generally pass the SNR=40 criterion at $z<0.5$, while they become looser at higher redshifts. This is a combined effect of the parameter selection seen on the top row of the same figure. The positive note is that fully disk dominated or bulge dominated systems, can be rather accurately modeled by single component model (see e.g. a discussion in \citep{2006MNRAS.371....2Allen2006}). Despite in this paper we will consider a similar SNR cut also for the 1-component analysis (see \S\ref{sec:1component}), in future we can test to reduce the SNR requirement for these simpler model and possibly reduce this incompleteness effect. 




\section{Training and Testing the {\rm GaLNet-BD}} \label{sec:floats}

As mentioned in \S\ref{sec:galnet} (see also Fig. \ref{fig:general}), the inputs of the GaLNet-BD are the 135 $\times$ 135 pixel galaxy images and the 51 $\times$ 51 pixel {\it PSF} images. The outputs are the 10 parameters, i.e. mag, $R_{\rm e}$, n, q, and PA, for the B and D components, which best predict the observed surface brightness distribution of each galaxy. Here below we detail the CNN training and testing of the B-D decomposition, using the simulated datasets introduced in \S\ref{sec:sim_gal}.

\subsection{Training and Validation} 
\label{sec:training}
The key to train any CNN is to minimize the loss function. Instead of traditional loss functions like mean square error (MSE) and mean absolute error (MAE), we choose "Huber" Loss function (\citep{friedman_2001}) with an “Adam” optimizer (\citep{kingma_ba_2014}). As discussed in Li+22, unlike MSE, "Huber" loss is less sensitive to outliers in the data by giving them smaller weights, hence allowing the CNN to quickly and efficiently converge by focusing on the low-scatter datapoints. Although, by definition, the MAE is also a little sensitive to outliers, it tends to give convergence problems as it does not efficiently weight the gradients in the loss function with the errors (for instance, it does not allow large gradients to converge in the presence of small errors).
This affects the overall convergence speed of the training process, sometimes even leading to convergence failures.
On the other hand, "Huber" loss has been proven to combine the advantages of MSE and MAE, providing us better accuracy and robust convergence. The "Huber" loss is defined as:
\begin{equation}
L_{\delta}(a) =\left\{
\begin{array}{lr}
             \frac{1}{2}(a)^2 , & |a|\le\delta  \\
             \delta \times (|a|-\frac{1}{2}\delta) , & ~~~{\rm elsewhere}
             \end{array}
\right.
\end{equation}
where $a$ is defined as 
$a=t-p$, in which $t$ is the label (real value) of the simulation and $p$ is the prediction value given by CNNs. While prediction deviation $|a|$ is smaller than $\delta$, the loss would be a square error; otherwise, the loss reduces to a linear function. After some trials on different $\delta$s, we have found that the CNN performs at the best when $\delta=0.001$.





\subsection{Statistical indicators} 
\label{sec:stat_meth}
To statistically assess the accuracy and precision of the predictions obtained from the GaLNet-BD, both in the training/validation phase and testing phase (see \S\ref{sec:results}), 
we adopt three diagnostics. 
~\\
1)    R squared ($R^{2}$), defined as:
\begin{equation}
     R^{2} = 1- \frac
     {\sum_{i}(p_{i}-t_{i})^{2}}
     {\sum_{i}(t_{i}-\bar{t})^{2}} ,
\end{equation}
where $t_{i}$ are the ground-truth values, $p_{i}$ are GaLNets’ predicted values, and $\bar{t}$ is the mean value of the $t_{i}$s. According to this definition, $R^{2}$ is 0 for no correlation (low accuracy) between ground-truth values and predicted values while 1 for the perfect correlation (high accuracy). This is a diagnostic that quantify how much the labeled input values and the CNN outputs are close to the 1-to-1 relation, i.e. the accuracy of the GaLNet-DB predictions.
~\\ 
2)    Normalized median absolute deviation (NMAD). 
We first define the relative bias as
\begin{equation}
     \Delta p =  \frac
     {p_{i}-t_{i}}
     {t_{i}},
     \label{eq:rel_bias}
\end{equation}

except for magnitude that are logarithmic quantities and for which the $\Delta p = p_{i}-t_{i}$. Then, the NMAD is defined as:
\begin{equation}
     {\rm NMAD} =  1.4826 \times {\rm median} ~(|\Delta p - {\rm median} ~(\Delta p)|).
     \label{eq:NMAD}
\end{equation}
This gives a measure of the overall scatter of the predicted values with respect to the 1-to-1 relation, i.e. the precision of the method.
~\\
3) Fraction of outliers.
This is defined as the fraction of discrepant estimates larger than $15\%$,  
using the condition $|\Delta p|>0.15$, similarly to what usually adopted for outliers in photometric redshift determination (see, e.g., \citep{Amaro2021+photz}). This gives a measure of the catastrophic predictions, which strongly deviate from the true values and can be driven by anomalous data, failures in the convergence of the CNN, etc.

The hyperparameters are optimized using the validation sample and found to reach the minimal of the loss after about a dozen of epochs. We also do not see signs of overfitting at later epochs.

After training, we use the test sample (see \S\ref{sec:sim_gal}) to check the GaLNets' performances and compare the ground-truth values of each parameter used to simulate the galaxies and the predicted values of disk components
and bulge components.

\begin{deluxetable}{lllllll}[t]
\tablenum{2}
\setlength{\tabcolsep}{0.85mm}
\tablecaption{\\Statistical Properties of the Prediction }
\tablehead
{\colhead{Test} & \colhead{Component} & \colhead{$mag$}& \colhead{$R_{\rm e}$}&\colhead{$PA$}&\colhead{$q$}&\colhead{$n$} }
\startdata
\hline
${R^{2}}$\\
&Disk      & 0.8723    &0.9237     & 0.8654   &0.9292      &0.1019  \\
&Bulge     & 0.9475    &0.8854     & 0.8654   &0.6416      &0.9875  \\
\hline              
Outlier frac.
&Disk      & 0.1004    &0.0257     & 0.1139   &0.0082      &0.1918  \\
&Bulge     & 0.0835    &0.0206     & 0.1142   &0.0218      &0.0067  \\    
\hline              
NMAD
&Disk      & 0.1137    &0.1753     & 0.0599   &0.1155      &0.3035  \\
&Bulge     & 0.1042    &0.1726     & 0.0599   &0.0863      &0.0537   \\ 
\enddata
\tablecomments{Statistical properties of the prediction on simulated testing data. From top to bottom we show $R^{2}$; the fraction of outliers and the NMAD for the magnitude $mag$; effective radius $R_{\rm e}$; position angle $PA$; axis ratio $q$ and S\'ersic index $n$. Generally the prediction of bulge component are better than those of Disk. }
\label{tab:stat}

\end{deluxetable}

\section{Test Sample Results and Discussion}
\label{sec:results}
In this section we discuss in details the GaLNet-BD performance 
over the test sample. As we have currently no real data at our disposal, this is the only sample we can use to benchmark the performance of the tool for future applications on observations. Although idealized, the simulated sample contains a rather high level of observational details in terms of seeing, noise, background and parameter distribution. Furthermore the 2-component models can capture most of the physical properties of real galaxies (see e.g. \citep{Gao_2017}), although the caveat in order is that presence of substructures in real galaxies  (like bars and spiral arms) can impact the correct parameter estimates in real applications (see e.g. \citep{2019ApJS..244...34Gao19}, \citep{2022ApJ...929..152L}).
Despite this latest limitation, we believe the simulated data adopted here is a fair knowledge base to train our tools for future 
CSST data (see 
e.g. Li+22 for a similar application to KiDS galaxies). 

We conclude this section by remarking that, regardless the specific application to CSST we discuss in this paper, the procedure illustrated above can be easily generalised to any other space instrument, provided that an accurate knowledge of the ``background'' and the {\it PSF} of the typical observations are available (see again \ref{sec:ccst_simu}, and also Li+22). 

\begin{figure}
\centering
\includegraphics[width=8.4cm]{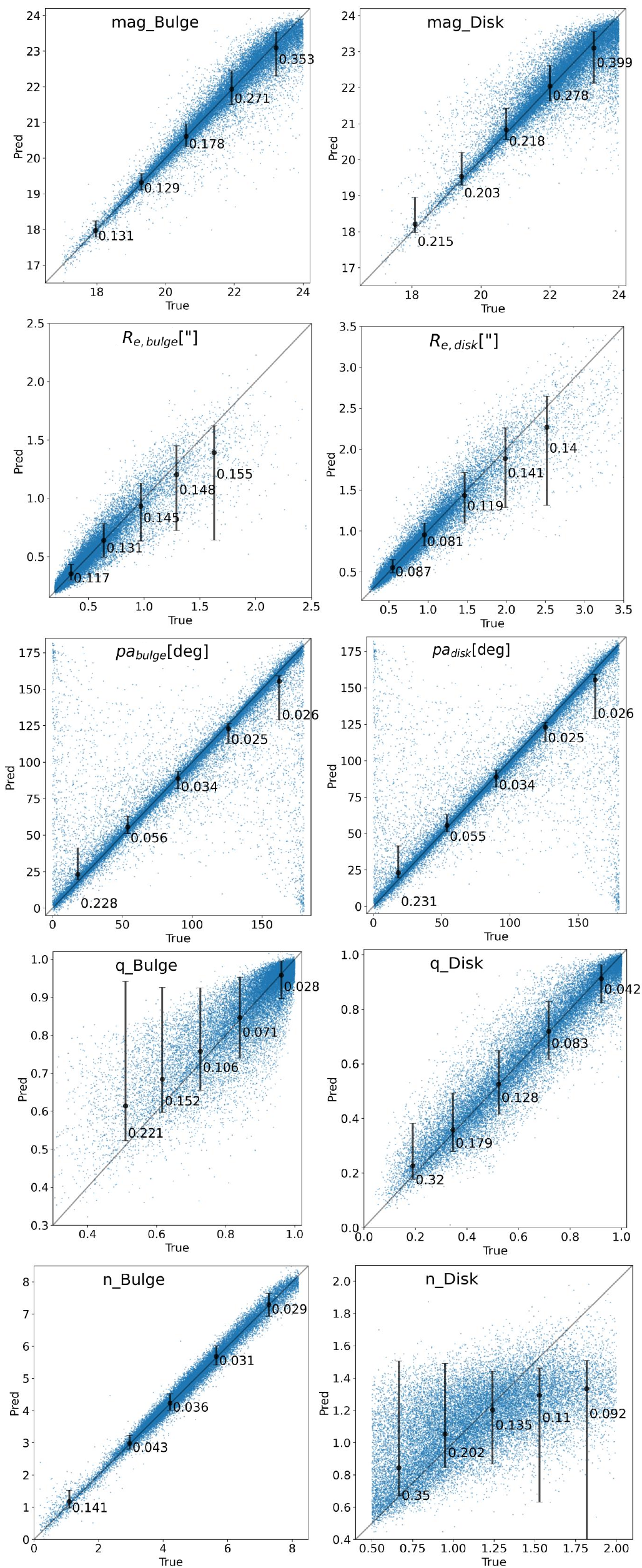}
\caption{Comparison between the true value and the predicted value of two components on simulated data. In each panel, the horizontal axes are the true values and the vertical axes are the predictions from GaLNets. 
The error bars are the absolute mean errors in each bin, while labels report the absolute errors for mag and relative errors for others. 
\label{fig:pred_linear}}
\end{figure}

\subsection{GaLNet-BD performances}
\label{sec:perf_linear}
The diagnostics obtained for all the 10 Bulge and Disk parameters are listed in Table \ref{tab:stat}, while the predictions (output) of the GaLNET-BD versus the ground truth values (input), for the 50k galaxies of test sample, are shown in Fig. \ref{fig:pred_linear}.
Looking at
this figure and the $R^2$ in Table \ref{tab:stat}, we find a very good accuracy ($R^2$) for the $mag$, $R_{\rm e}$, and PA for both disk and bulge, although for the PA there is a rather large fraction of outliers around 0/180 deg, mainly driven by the round systems ($q\sim1$), for which the PA is rather uncertain. For the three quantities, we find similar scatters (NMAD) and outlier fractions for both Bulges and Disks (from Table \ref{tab:stat}). These show that $mag$, $R_{\rm e}$, and PA are rather robustly constrained.

\begin{figure*}[t]
\centering
\includegraphics[width=18cm]{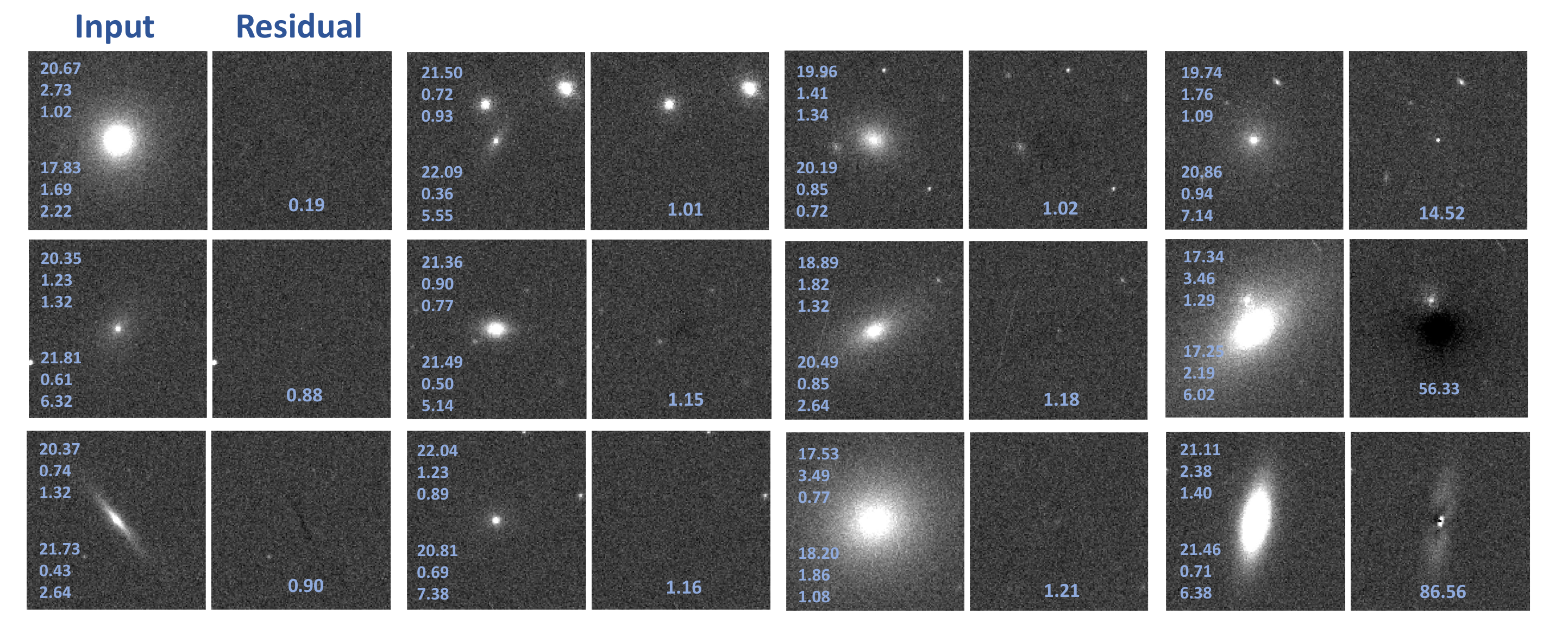}
\caption{Residual maps, obtained from simulated ``galaxy'' images (left panels), after subtracting the reconstructed S\'ersic models using GaLNets-BD trained on normal ``Residual images'' (right panels)  -- see \S\ref{sec:perf_linear}. At the left side of each input image, we report the $mag$, $R_{\rm e}$ and  and $n$ of disk and bulge.  We also report the a posteriori reduced $\widetilde{\mathcal{X}}^{2}$ at the bottom of each residual image. Images are ordered by $\chi^2$ (top to bottom), with the left column presenting cases with $\tilde{\chi}^2<1$, the two central columns the cases with $1<\tilde{\chi}^2<2$ and the right column $\tilde{\chi}^2>>2$.  
\label{fig:residuals}}
\end{figure*}

Moving to the other parameters, the S\'ersic index of the bulge, $n_{\rm bulge}$, is rather well constrained, showing a high $R^2$ and small outlier fraction. 
On the other hand, the $n$-index of the disk, $n_{\rm disk}$ is poorly predicted ($R^2=0.10$, see Table \ref{tab:stat}). 
We see an opposite situation for the axis ratio. The $q$ of the bulge, $q_{\rm, bulge}$, is poorly predicted ($R^2=0.64$) and show a rather large fraction of outliers, while the one of the disk, $q_{\rm disk}$, shows a much higher accuracy ($R^2=0.93$). This is also well understood as the bulge is embedded in the disk and its axis ration gets easily mixed with the disk, while the disk geometry is better defined toward the edge of the galaxy.


The most strident result is the bad performance of the $n_{\rm disk}$ with respect to the tighter constraints on the $n_{\rm bulge}$, although not unexpected. 
In fact, in the galaxy center, the $n_{\rm bulge}$ is generally dominant over the one of the disk, and 
the CNN tends to associate the peak in the central, high SNR regions of a galaxy image, to the dominant component (the bulge). On the other hand, the smoother peak of the lower $n$-index ($<2$) disks is generally embedded in the bulge intensity. Here, the only way for the GaLNet-DB to constrain the $n_{\rm disk}$ 
is to guess it, together the other disk parameters, from the outer galaxy regions, where the disk dominates.
Indeed, as also discussed in Li+22, the GaLNets seem to learn how to predict the parameters from the light density gradients. For this reason, we can also see that the $R_{\rm e}$ of the disk, $R_{\rm e,disk}$, is better predicted than the one of the bulge, $R_{\rm e,bulge}$, because disks dominate in the lower SNR outer regions and the GaLNet-BD can correctly recover the main parameters connected to the surface brightness gradient far from the center (i.e. effective radius and total luminosity, not the $n$-index as discussed above). Going toward the center, though, the gradients of the bulge component is strongly affected by the disk density profile around the $R_{\rm e,bulge}$, and, because of that, this latter parameter is slightly worse constrained than the $R_{\rm e,disk}$ (see Fig. \ref{fig:pred_linear} and $R^2$ in Table \ref{tab:stat}).

Finally, in Fig. \ref{fig:residuals}, we show a sample of data/residual images.
For each example, on the left, there is the galaxy cutout (labeled as ``Input'') used as input for the GaLNet-BD, where we also report the ``true'' parameters. 
On the right (labeled as ``Residual''), the residual image after the predicted 2D model from the CNN has been subtracted, with (at the bottom) the 
the reduced $\tilde \chi^2$ defined following Li+22:
\begin{equation}
\tilde{\chi}^2=\sum_{i\neq0} \frac{(f_i - m_i)^2}{\sigma_{\rm bkg}^2}/{\rm dof} 
\label{eq:chi2}
\end{equation}
where $f_i$ are the observed pixel fluxes of the galaxy within the effective radius, $m_i$ the model values in each pixel, the $\sigma_{\rm bkg}^2$ the background noise, and  dof $=$ (N. pixels $-$ N. fit parameters). Note that, in Eq. \ref{eq:chi2} we exclude the central pixel ($i=0$) because it is too sensitive to small variation of the parameters (especially the S\'ersic index), during the convolution with the {\it PSF} used to reconstruct the modeled galaxy. This could cause strong deviations from the true ``observed'' fluxes at $i=0$, hence artificially degrading the overall $\chi^2$.
The Fig. \ref{fig:residuals} shows three different groups of best-fit ``goodness'', the very good ones ($\tilde{\chi}^2<1$), the mid quality ones ($1<\tilde{\chi}^2<2$), and the bad ones ($\tilde{\chi}^2>>2$).   

From the figure we can see that the performances of the GaLNet-BD is generally good both for isolated galaxies and for more crowded situation, where there are close systems. Major failures occur if the galaxy has a compact bulge and disk (e.g. high redshift) in crowded area, or for very bright systems ($r>18$). We believe this latter case is partially due to the smaller number of galaxies present in the training sample for the bight magnitude range (see e.g. Fig. \ref{fig:z-mag}). We discuss the impact of the distribution of priors for the training sample in Appendix \ref{sec:app}.

\begin{figure}
\hspace{-0.8cm}
\includegraphics[width=9.5cm]{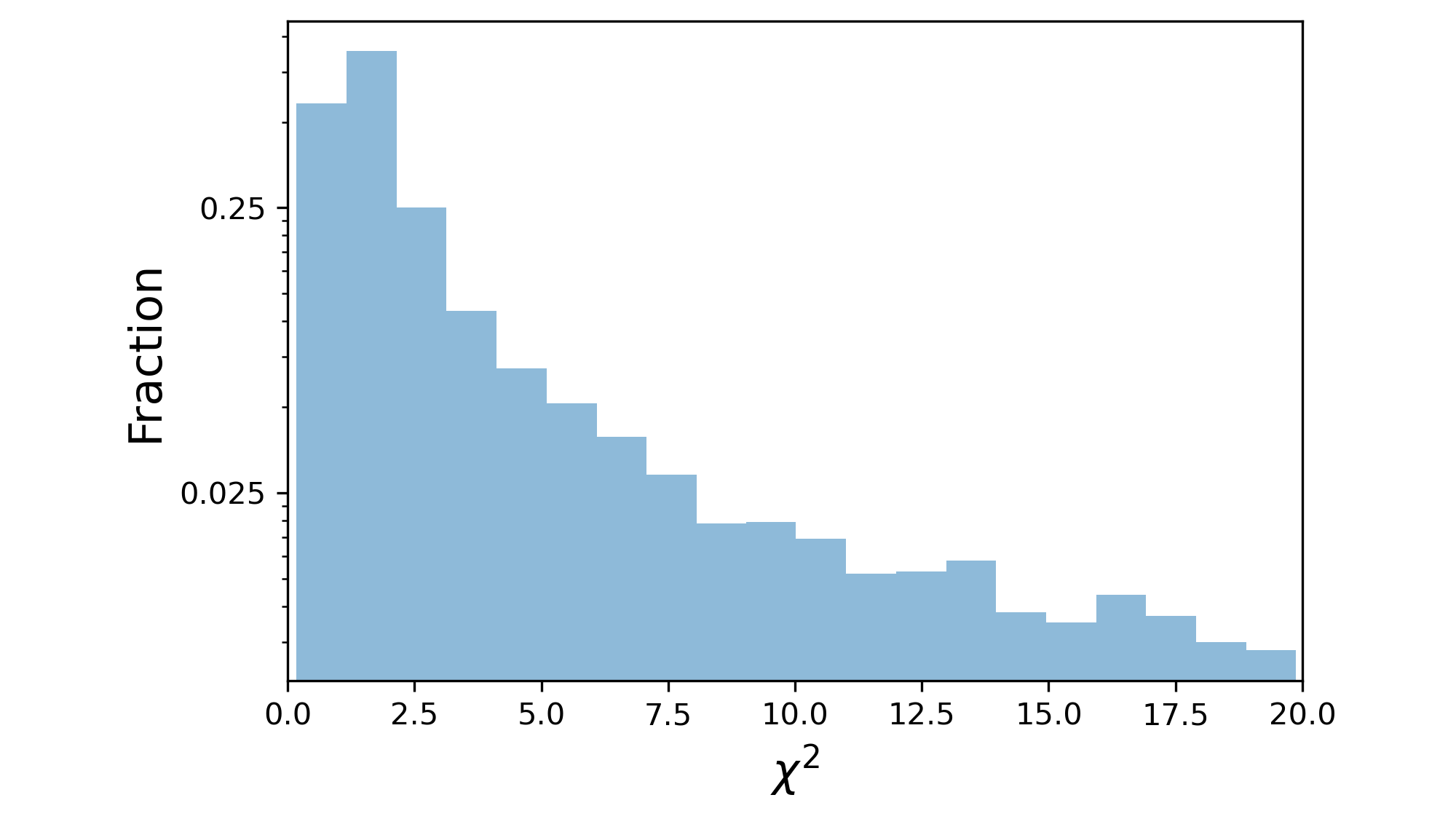}
\caption{Distribution of the $\chi^2$, as defined in Eq. \ref{eq:chi2}, for the GaLNet-DB predictions. The peak at $\chi^2<2$ contains almost the $\sim$60\%.
\label{fig:chi2_lin_log}}
\end{figure}

\begin{figure}
\hspace{-0.8cm}
\includegraphics[width=10cm]{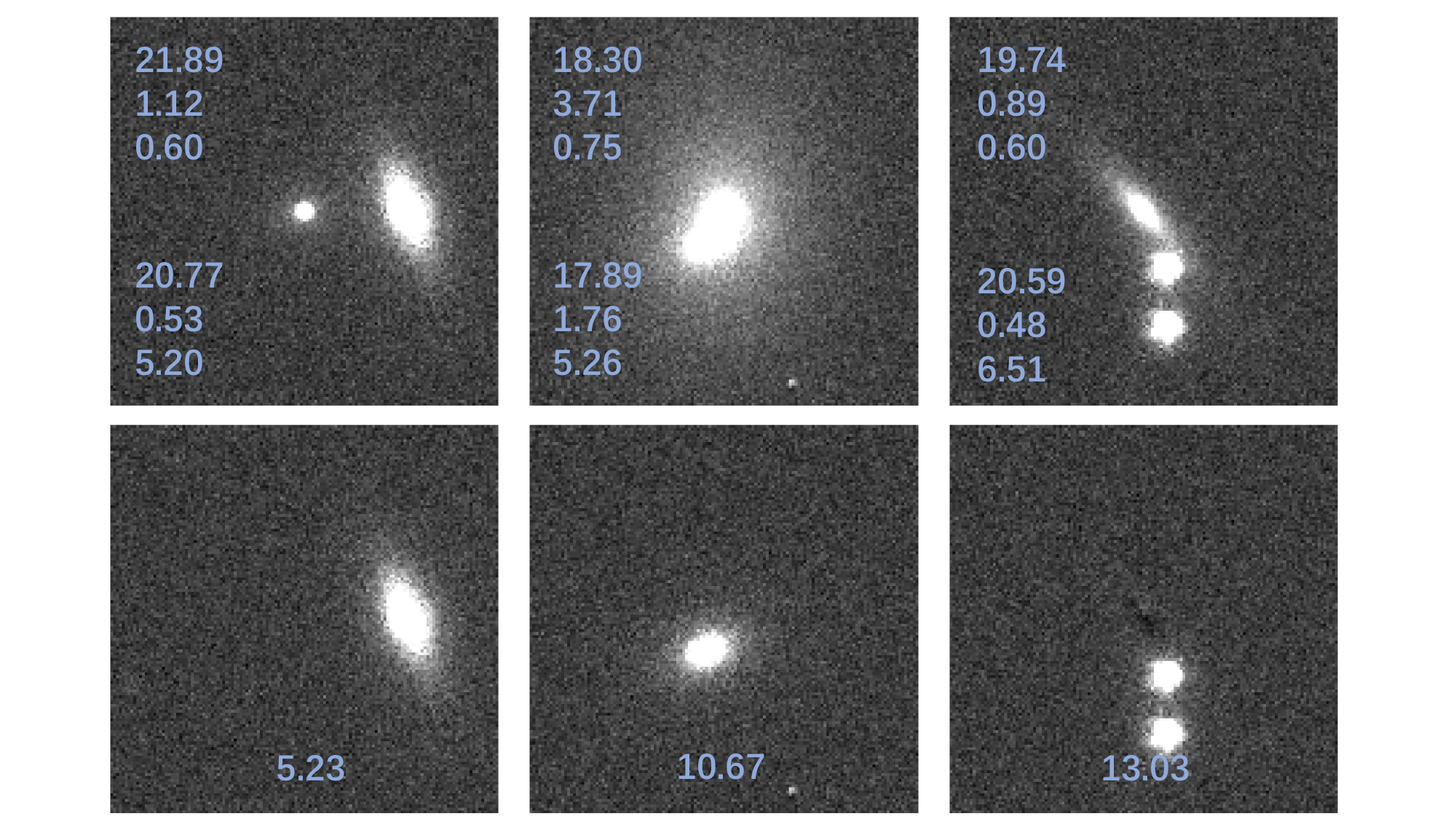}
\caption{Some examples of residual map exhibiting "good'' fitting but bad $\tilde{\chi}^2$ due to the presence of a close or blended sources. Top: mock galaxies in the center of the field; bottom: residual images. As for Fig. \ref{fig:residuals}, the left side of each residual image, we report the $mag$, $R_{\rm e}$ and  and $n$ of disk and bulge.  We also report the a posteriori reduced $\widetilde{\mathcal{X}}^{2}$at the bottom of each residual image. 
\label{fig:goodfit_badchi2}}
\end{figure}

In Fig. \ref{fig:chi2_lin_log}, we finally show the distribution of the $\chi^2$ for the full test sample. We can see here that $\sim 60\%$ of the sample shows a good fit ($\tilde{\chi}^2<2$), which represents a reasonably high fraction. We remark here that this might be a lower limit as we are not subtracting the cases where the model is good even in presence of a blended source, which can still contribute to the residual image in the $\chi^2$ calculation. To have a measure of these possible ``contaminants'', that can even make the prediction to fail (see e.g. Fig. \ref{fig:residuals}), we have estimates the numbers of background images with RMS larger than the median value of the majority of them (RMS$\sim$32 counts) to be of the order of 15\%. This means that very likely a significant fraction of the 40\% having $\tilde{\chi}^2>2$ might have still a rather ``good'' fit and residual map. We show some of these examples in Fig. \ref{fig:goodfit_badchi2}.



As a final note, the $\chi^2$ is not a measure of the accuracy of the predictions, but, rather, of the ability to reproduce the observed surface brightness distribution of the galaxies. Due to the degeneracies among the parameters, the predicted target values can deviate from the ground truth, but yet combine to give a good ``fit'' to the data. This is possibly an intrinsic problem of the multi-parameter fitting that does not have a simple solution. One possibility will be to use a multi-band approach (see e.g. \citep{2014MNRAS.444.3603V}, \citep{2022A&A...664A..92H}), and trying to minimize the systematics on the S\'ersic parameters assuming that they do not change too much from one band to another. We will address this multi-band analyses in a future paper, where we foresee that using ``transfer learning'' from one band to the others will likely help break these degeneracies.  

\begin{figure*}[t]
\hspace{-0.5cm}
\includegraphics[width=20cm]{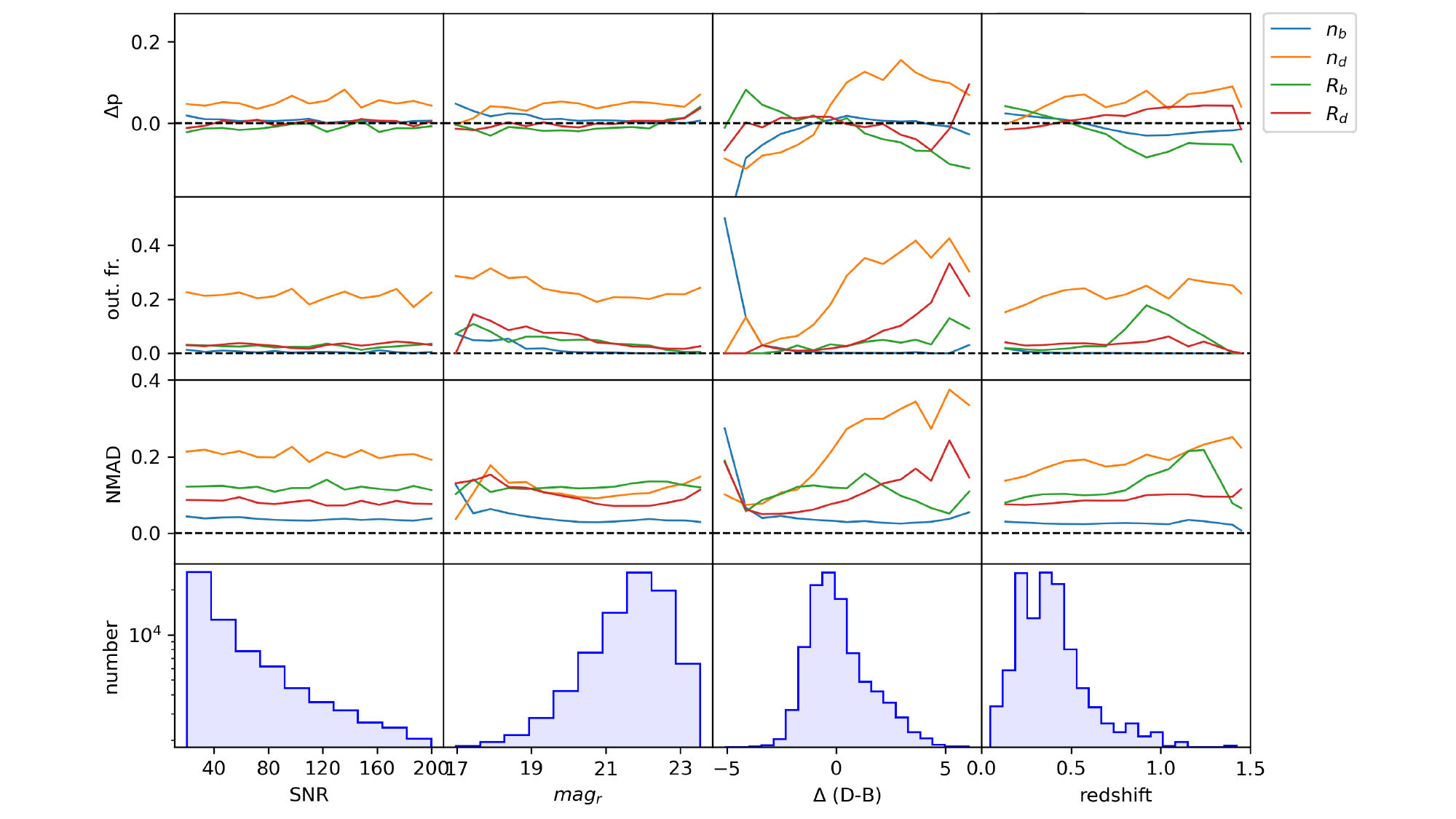}
\caption{Bias ($\Delta p$), Outlier fraction (out. fr.), and Scatter (NMAD) as functions of SNR, total magnitude, $\Delta$(D-B)$=mag_{\rm disk}-mag_{\rm bulge}$ in $r$-band, and redshift. In each panel, blue and orange lines are for S\'ersic index of bulge and disk, respectively, and green and red for effective radius of bulge and disk, respectively (see legenda). Last row at the bottom: the number distribution in the corresponding parameters. 
\label{fig:snr_mag_z}}
\end{figure*}

\subsection{Accuracy and Precision dependence on SNR, mag, D/B and redshift}
\label{sec:snr_mag_z}
To conclude this section we want to check the performances of the GaLNet-DB as a function of specific parameters that can differently impact the accuracy and precision of the predicted targets. In particular, we have seen that the high-SNR is a pre-requisite for accurate predictions, at the cost of a significant selection effect on the accessible parameter space (see Fig. \ref{fig:z-mag}). Hence, it is natural to ask 
what degradation of the accuracy and precision we need to expect for SNRs close to the lowest adopted limit and, consequently, to the faintest reachable magnitudes. 
A different piece of argument comes with redshift. Although the high$-z$ sample fully obeys the SNR requirements, a further complication of its analysis comes from the intrinsically small galaxy angular sizes, implying that most of the galaxy information is concentrated in a dozen of pixels around their centers. In this case, 
the performance of the GaLNet-DB can be ruled by the limited number of features, which is close to the number of parameters that the CNN aims at constraining, rather than the SNR. This is a situation equivalent to having a limited number of degrees of freedom in standard best-fitting techniques, which causes either high-degeneracy among the parameters or rather large uncertainties. 
Finally, we also consider the total magnitudes in $r$-band, and the Disk-to-Bulge, $\Delta$(D-B). This latter property is particular important to see how the relative mix of the two components might impact the ability of the \GAL-DB to infer their intrinsic structural parameters.

In Fig. \ref{fig:snr_mag_z} we give an impression of the accuracy and precision of the predictions using three indicators. First, the relative accuracy 
is measured by the relative bias, $\Delta p$, defined in Eq. \ref{eq:rel_bias}. Then, the outlier fraction and the scatter as measured by the NMAD.
They are all derived for the main structural parameters of the Bulges and Disks, i.e. the $n$-indexes and $R_{\rm e}$s.
Finally, in the bottom row of the same figure, we report the distributions of the test sample parameters, that mirror the ones of the training sample, by construction.
From left to right we notice that: 
\\~\\
1) The $\Delta p$, outlier fraction and NMAD, show almost no variation with the SNR, meaning that the limit of ${\rm SNR}=40$ is well justified. For the bulge and disk $R_{\rm e}$ and for the bulge S\'ersic index the bias and outlier fractions are almost absent and the NMAD show a rather smal scatter, only $n_{\rm disk}$ shows 
a rather systematic 
offset of $\sim$5\%, which is yet within the scatter (NMAD$\sim 0.2$). The constancy of the offset and scatter suggest the systematics of the $n_{\rm disk}$ are independent of the SNR.


2) The behaviour of the statistical indicators as a function of the ($r$-band) luminosity mirrors the one of the SNR in the left column. The $n_{\rm disk}$ is the quantity that shows the largest systematic offset and outlier fraction. The NMAD in the other hand is similar to the one of the other quantities. From these plots we also conclude that there is not significant variation on the overall accuracy and precision of the predictions as a function of the luminosity. 
\\~\\
3) On the other hand, there is a clear dependence of the all indicators on 
the $\Delta$(D-B). When disks dominate (i.e., $\Delta$(D-B)$\lsim0$), almost all indicators are reasonably small ($|\Delta p|<0.1$, out. frac. $<0.1$ and NMAD$<0.1$), except for very small $\Delta$(D-B) ($<-3$) where the training/test samples are underrepresented. For $\Delta$(D-B)$\gsim0$ the disk parameters start to degrade (especially the outlier fraction and NMAD), with the disk S\'ersic index being the worst affected parameter. This is consistent with what seen in Fig. \ref{fig:pred_linear} and Table \ref{tab:stat}. This suggests that the real driver of the systematic of the disk parameter is the presence of a dominant disk. We also note that there is a trend of the $R_{\rm e,bulge}$ to degrade in accuracy (i.e. larger negative $\Delta p$) for dominant bulges ($\Delta$(D-B)$\gsim0$), which seems counter-intuitive, but that we can track to the decreased sample size (see histogram at the bottom), due to the lower density of bulge dominated systems at higher-$z$ (see below).

4) There seems to be a weak trend of the three indicators toward a degradation at higher redshift for 
$n_{\rm disk}$ and  $R_{\rm e,bulge}$.
As we have seen in Table \ref{tab:stat} and discussed in \ref{sec:perf_linear}, these are the two quantities that are recovered with lower accuracy in general, and moving toward higher redshift, these quantities result to be less accurate and more noisy. 
This is mainly due to the small angular-size of galaxies that reduce the number of pixels reaching enough SNR in the outskirt to perform an accurate analysis of the density profiles.
For the bulge quantities, at high-$z$, there is the additional problem (see above) of the poorer training sample due to the lower numerical density of dominant bulges.
We finally notice that for $z<1$, all the indicators are stably at the same level of the average values found as a function of the SNR (in the left panels), i.e. $\Delta p\sim$5\%, NMAD$\sim$0.15 or smaller and oulier fractions $<$5\% except for the $n_{\rm disk}$ which is $\sim 25\%$.
This shows that $z\leq1$ is possibly a conservative upper limit for the B-D analysis, with 1-epoch data. Likely, with deeper images, the SNR of these systems will reach higher levels for further away regions, hence increasing the number of pixels to be used as features from the CNN. 
Along the same line of arguments, we can also expect that limiting the number of parameters to constrain, we can possibly push this limit further ahead in redshift. E.g. as not all galaxies are multi-component systems, but can often be well approximated by a single dominant component,
we can check whether using a 1-S\'ersic GaLNet we can reach high accuracies and precision for higher-$z$ galaxies. With this aim, in the next section we will assume a population of 1-component systems still represented by a general S\'ersic profile with a wide range of parameters spanning from disks to spheroidal galaxies.
\begin{figure}
\vspace{-0.cm}
\hspace{-0.5cm}
\includegraphics[width=8cm]{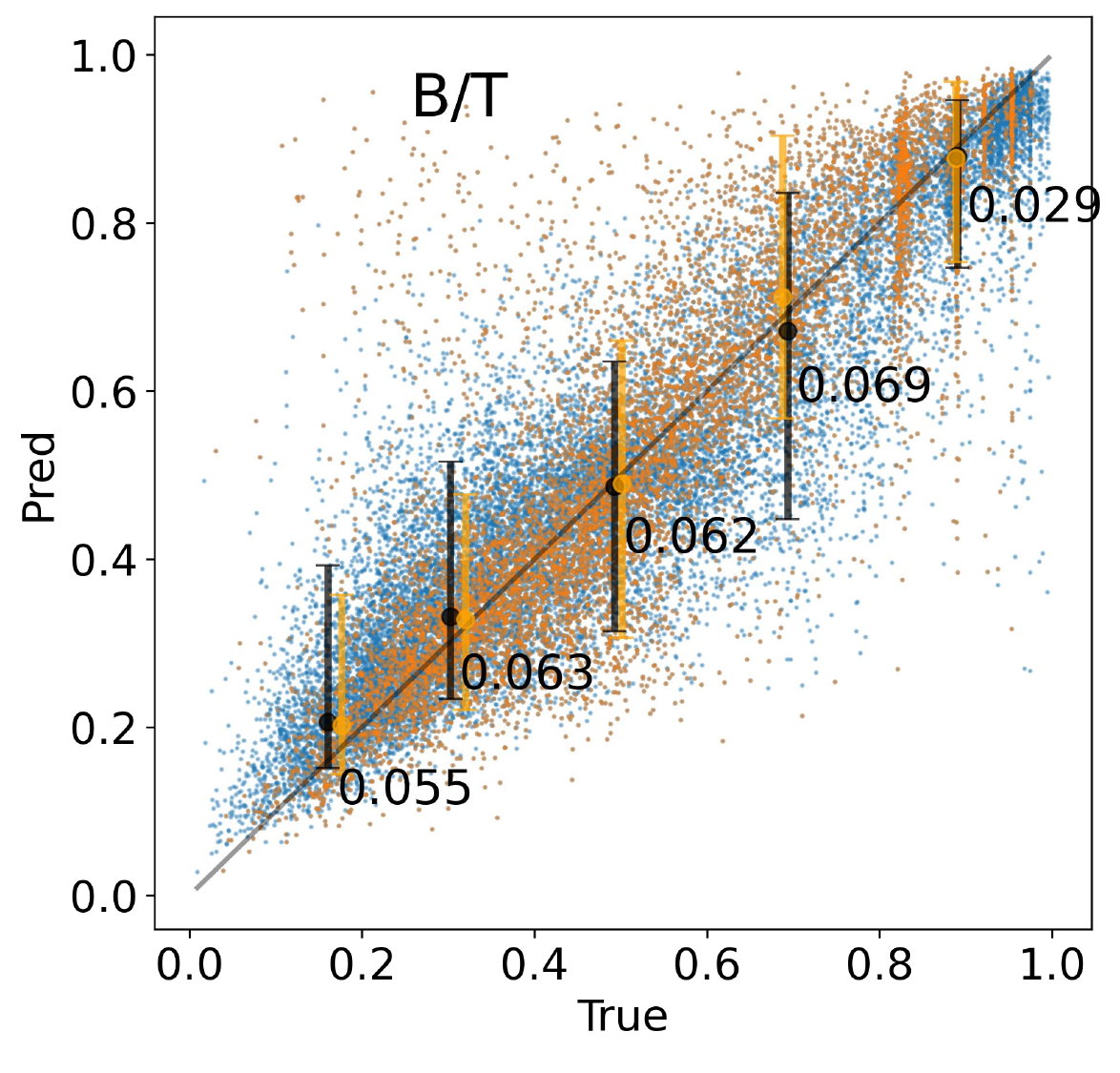}
\caption{Comparison between the true B/T and the predicted B/T of two components on the full sample of mock CSST galaxies (blue dots) and the ``bright'' sample with $mag_{\rm disk}$ and $mag_{\rm bulge}$ being selected to be $r<22$ (orange points). The horizontal axes are the true values and the vertical axes are the predictions from GaLNets. 
The error bars are the normalized median absolute deviation (NMAD) in each bin. The four stripes of galaxies at B/T$>0.8$ come from some peaks in the CosmoDC2 distributions. 
\label{fig:BT}}
\end{figure}

\subsection{On the B/T prediction} 
A clear result of the B-D decomposition, performed in the previous section, is that the best constrained targets are the bulge and disk magnitudes (mean $R^ \sim 0.9$). Typical precisions of $mag_{\rm bulge}$ and $mag_{\rm disk}$ are of the order of 15\%, although we also observe a significant degradation at magnitudes $r>22$. Looking at these quantity separately, though, does not give a sense of the overall accuracy of the bulge-to-total ratio, B/T, which is a standard proxy of the galaxy morphology (\citep{1986ApJ...302..564Simien_DeVauc}, \citep{Graham_B/T}) and correlate it with other physical parameters (e.g. \citep{2013ARA&A..51..511Kormendy_Ho_13}, \citep{10.1093/mnras/stu594Bluck_B/T}). Hence, here we want to explicitly quantify the accuracy and precision of the B/T derived by our B-D decomposition. In Fig. \ref{fig:BT} we show the predicted vs. true B/T values. The former are obtained from the ratio the predicted magnitude of the bulge and total predicted magnitudes, while the latters are the same input quantities from CosmoDC2. From the data in Fig. \ref{fig:BT}, we have estimated an $R^2$ of 0.80, an NMAD of 0.06 and outlier fraction of 0.05. Note that being the B/T a quantity smaller than one, by definition, we have re-defined the $\Delta p= (p_{i}-t_{i})/(1-{t_{i}})$ in Eqs. \ref{eq:rel_bias} and \ref{eq:NMAD}, similarly to what is usually done for galaxy redshifts (e.g. \citep{Amaro2021+photz}).  
The $R^2$ we obtain is worse than the ones found for the bulge and disk magnitudes separately, suggesting a lower overall accuracy of the B/T. This might come from the tails of the faint bulges and disks seen in Fig. \ref{fig:pred_linear} at $r>22$, which is propagated in the B/T plot above.
To check that, in the same figure we also plot (orange points) the B/T predictions of 
a ``bright'' sample defined as the galaxies 
having $mag_{\rm disk}$ and $mag_{\rm bulge}$ smaller than $r=22$ (orange points and errorbars in Fig. \ref{fig:BT}). In this case we obtain a $R^2=$0.83, NMAD=0.06 and outlier fraction of 0.06, which are almost equivalent to the full sample. This means that the B/T parameter is very sensitive to the intrinsic scatter of the magnitudes (regardless how small) of the two components, which is turned into a low accuracy. Looking at Fig. \ref{fig:BT}, we also notice that the maximum of the scatter (and outlier fraction) is concentrated in the interval $\rm 0.2<B/T<0.8$ while at small and large B/T, where either the disks, or the bulges dominate respectively, the scatter is reduced. This suggests that the uncertainties are larger where there is a coexistence of the two components and are reduced in presence of a dominant component. In this latter case though, we notice that being the disks more poorly constrained (see Table \ref{tab:stat}), at B/T$<0.2$ there are some clear systematics, which are reduced for the ``bright'' sample (see orange mean datapoint at B/T$<$0.2). We notice that a systematic overprediction of the B/T was found also by the CNN presented in \citet{2021MNRAS.506.3313G}, although the two results cannot be directly compared as their analysis is based on a completely different approach (no B-D decomposition) and data (ground based, bright $r<18$ galaxies).

\subsection{One-component galaxy systems} \label{sec:1component}

The 2-S\'ersic profiles reproduce the majority of lenticular and late-type systems. However, most of the elliptical galaxies are characterized by a dominant spheroidal component, if one excludes the extended stellar haloes which are ubiquitously found around bright elliptical galaxies at very faint surface brightness levels (\citep{2016ApJ...820...42I}, \citep{2020arXiv200713874D}).
These latter are generally detected at low redshift and are possibly difficult to be recognised at $z>0.5$ (e.g. \citep{1994Natur.370..441S}; \citep{2004MNRAS.352L...6Z}; \citep{2012arXiv1204.3082B}, but see, e.g., \citep{10.1093/mnras/stt232,2022arXiv220905519G}).
Hence, there is a large variety of galaxies that are well described by a 1-S\'ersic profile. For these systems, due to the lower complexity and absence of strong degeneracies introduced by the superposition of multi-components, we can reasonably expect to obtain robust GaLNets predictions even at $z>1$, which somehow sets an upper limit for the GaLNet-BD on 1-epoch CSST data
 as seen in \S\ref{sec:snr_mag_z}.
In the same section, 
we have also discussed that the major limitation imposed by the high$-z$, even for space observations, 
is the small angular size of galaxies, that reduces the number of useful pixel data to constrain a large number of parameters. By limiting ourselves to 1-S\'ersic profiles we reduce the number of the parameters by a factor of two, hence leaving a larger number d.o.f. for our models.


\begin{figure}
\vspace{-0.cm}
\hspace{-0.5cm}
\includegraphics[width=9cm]{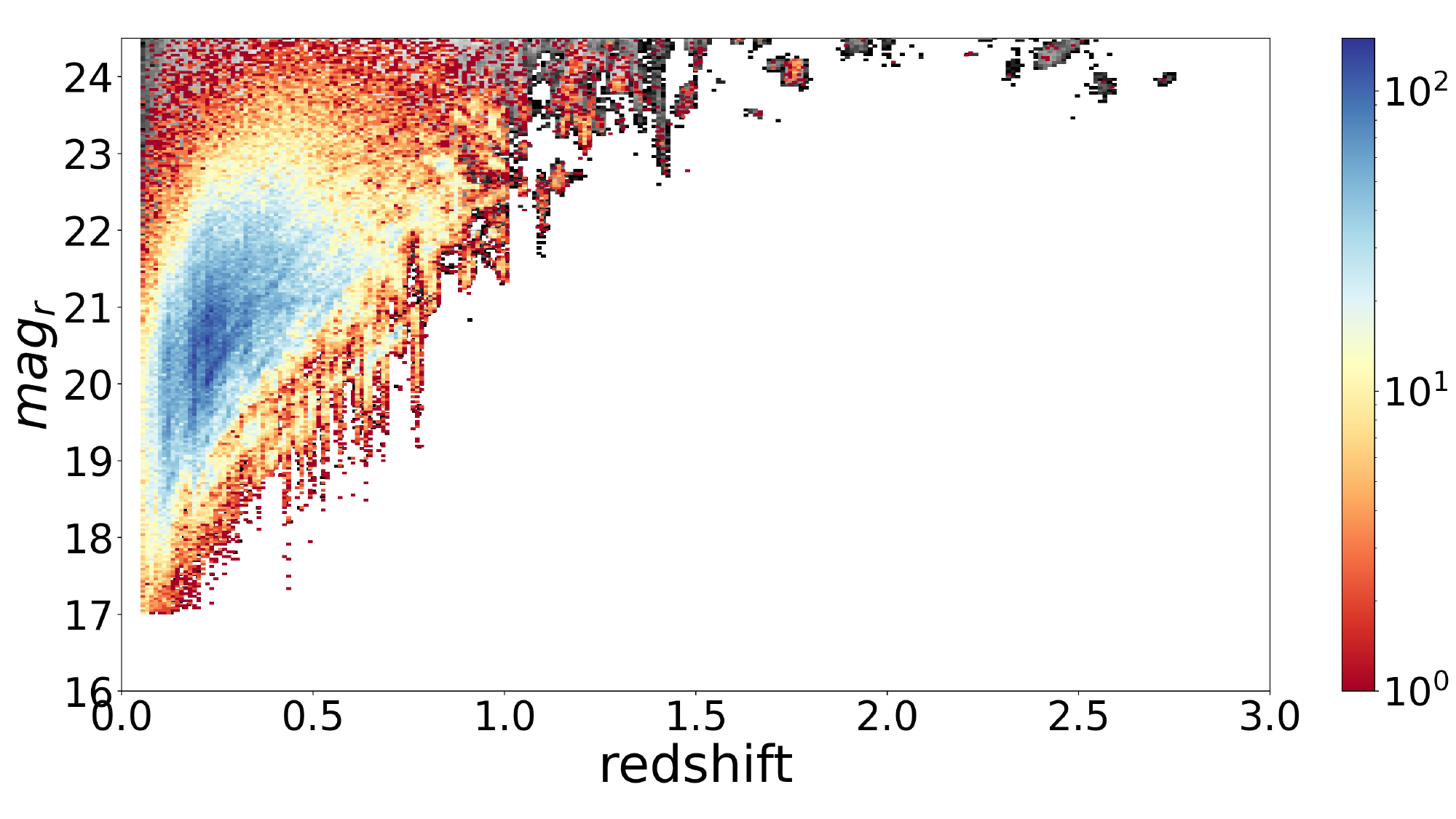}
\caption{Redshift versus total magnitude before (gray points) and after (color + bar) the SNR cut,
for the 1-S\'ersic model run. 
At a fixed SNR, the 1-S\'ersic models result generally fainter than the 2-S\'ersic ones. This is due to the fact that the disk component generally contribute with a shallower profile that dilute the SNR inside the $R_{\rm e}$ as defined in in \S\ref{sec:SNR_cut}. Compared to the 2-S\'ersic in Fig. \ref{fig:z-mag}, here galaxies pass the ${\rm SNR}=40$ criterion down to $mag_{r}>23.5$ and redshift $z>1.5$, although they are too sparse at $z\gsim1.75$ and we will retain only galaxy below this redshift limit. 
\label{fig:mag_redshift}}
\end{figure}

In this test, we use the same method and CNN architecture as adopted in the first GaLNet work (Li+22). In particular, we use the GaLNet-2 and train it over the CSST mock observations and local {\it PSF} (as in \S\ref{sec:sim_gal}). For this purpose, we randomly collect 200\,000 galaxies
from the 1-S\'ersic galaxy catalog provided by the CosmoDC2.
Even in this case, though, we needed to override the $n=4$ set in CosmoDC2 for the 1-S\'ersic models, and instead use the a more realistic log-normal $n$-index distribution as from Li+22. 
After having produced the {\it PSF} convolved models, with Poisson noise, and added these to the ``Background'' images as described in \S\ref{sec:sim_gal}, we impose the condition of SNR\textgreater40. Once again, this produces an alteration of the final parameter distribution, but less severe than the 2-component model, as shown in Fig. \ref{fig:mag_redshift} for the magnitude distribution vs. redshift of 200\,000 mock galaxies as compared with the original CosmoDC2 catalog. Here we see that there is no sharp drop of magnitudes after $r\sim23$, while there is a rather large, albeit patchy, population of galaxies at $z>1.5$.
The non uniform distribution is possibly due to the volume and resolution of CosmoDC2 simulation, which looses details of the field galaxies and picks ``Large Structure'' populations at high redshift, as the ones shown around $z=2$ and $z=2.5$ in Fig. \ref{fig:mag_redshift}. We finally decide to avoid the sparse sample at $z\gsim1.75$ and us this latter as upper limit of our analysis.


\begin{figure}
\centering
\includegraphics[width=8.2cm]{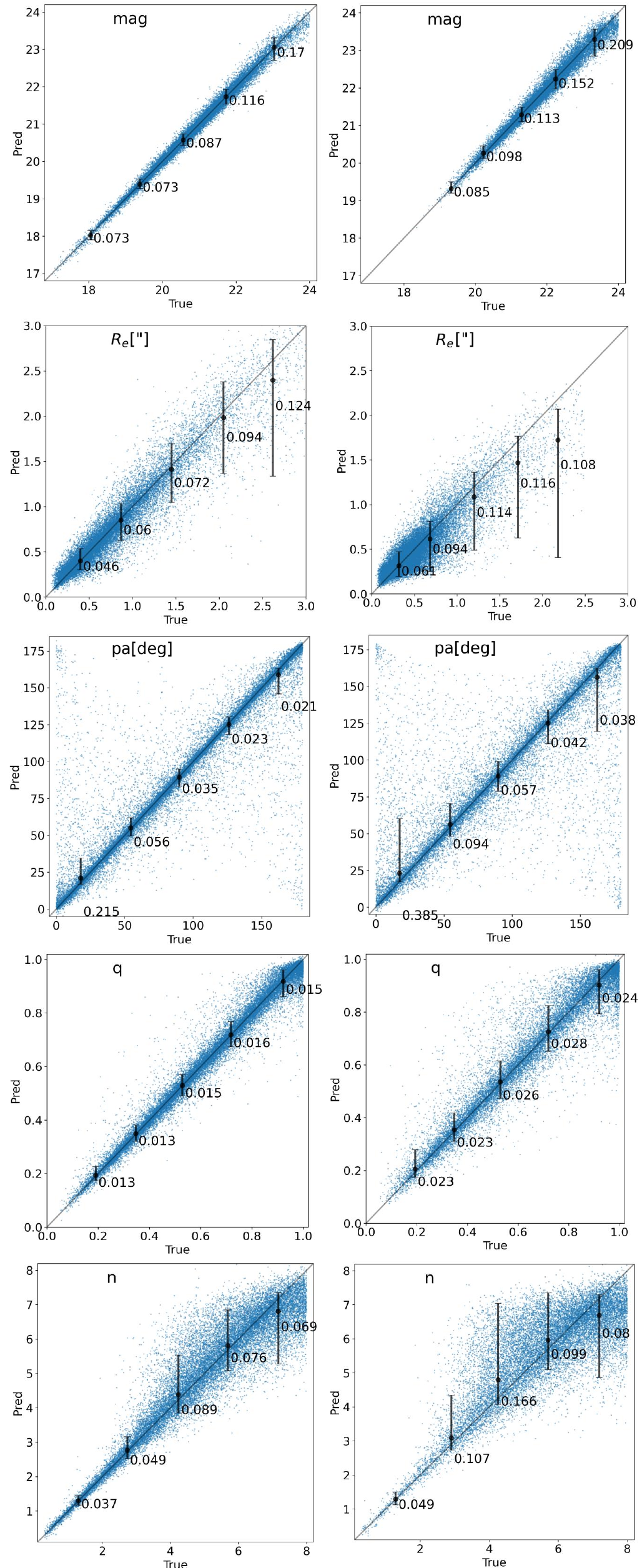}
\caption{Comparison between the true value and the predicted value of low redshift samples ($z<0.5$, left panel) and high redshift samples ($z>0.5$, right panel). In each panel, the horizontal axes are the true values and the vertical axes are the predictions from GaLNets. 
The error bars are the normalized median absolute deviation (NMAD) in each bin, while labels report the absolute errors for mag and relative errors for others.
\label{fig:pred_1S}}
\end{figure}


The training and testing of the GaLNet-2 both follow the same steps as described in \S\ref{sec:floats}.
Furthermore, in order to fully evaluate the model performance on high redshift galaxies, besides the test sample of 50\,000 galaxies randomly taken at all redshifts (full-$z$, hereafter), we specifically selected 25\,000 galaxies with redshift $>0.5$ as a further high-$z$ test sample (high-$z$, hereafter). 

\begin{figure}
\hspace{-0.7cm}
\includegraphics[width=9cm]{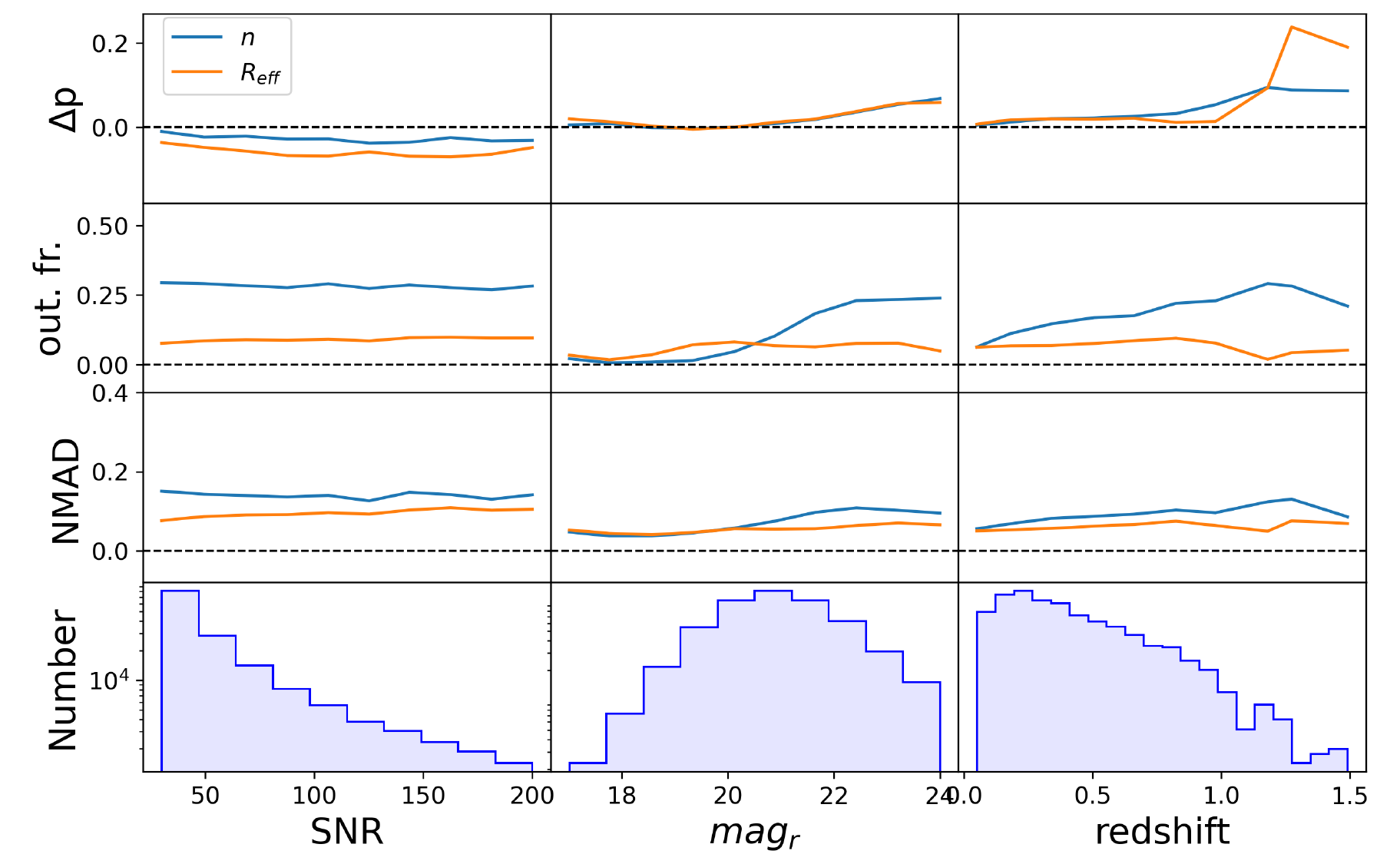}
\caption{Accuracy and precision for 1-component systems. Bias ($\Delta p$), Outlier fraction (out. fr.),  and Scatter (NMAD) as functions of spec-z, photo-z, and magnitudes in 20 bins for the ``Log images'' sample. In each panel, blue line is for S\'ersic index and orange is for effective radius. In the last row we also present the number distribution in the corresponding parameter space. The y-axes scales are the same as the Fig. \ref{fig:snr_mag_z} for comparison.
\label{fig:snr_mag_z_1}}
\end{figure}

The final results on the two test samples are reported in Fig. \ref{fig:pred_1S}, where we plot the predicted parameters vs. ground truth values. In Table \ref{tab:stat_1}, we report the statistical indicators for the predicted targets broken in the low-$z$ ($z<0.5$) and high-$z$ ($z>0.5$) samples. For the low-$z$ sample, we can see a general good accuracy of the main galaxy parameters ($mag$, $R_{\rm e}$ and $n$), with small systematics only at higher $n$. The accuracy is degraded for the high-$z$ sample, especially for a certain tendency to underestimate the effective radii at $R_e>1''$ and to overestimate the S\'ersic index at $n=4-6$ and understimate at larger $n$. This is reflected in Table \ref{tab:stat_1}, where we clearly see that $mag$ and $q$ are rather accurately reproduced at all redshifts, while the indicators all degrade for the higher redshift sample. 
This is also shown in more details in Fig. \ref{fig:snr_mag_z_1}, which contains the statistical indicators as a function of SNR, $mag$, and redshift.

\begin{deluxetable}{lllllll}
\tablenum{3}
\setlength{\tabcolsep}{0.85mm}{
\tablecaption{\\Statistical Properties of the 1-S\'ersic-Prediction }
\tablehead
{\colhead{Test} &{Components} & \colhead{mag}& \colhead{$R_{\rm e}$}&\colhead{PA}&\colhead{q}&\colhead{n} }
\startdata
\hline
$R^{2}$ 
&low redshift&0.9981    &0.8910     & 0.8415   &0.9776      &0.8894  \\
&high redshift &0.9782  &0.7641     & 0.7410   &0.9332      &0.6718  \\
\hline              
Outlier Frac. 
&low redshift           &0.0005    &0.0707     & 0.1718   &0.0013      &0.1053  \\
&high redshift          &0.0015    &0.0910     & 0.2485   &0.0100      &0.1875  \\   
\hline              
 NMAD\
&low redshift            &0.0095    &0.0536     & 0.0387   &0.0134     &0.0661  \\
&high redshift           &0.0144    &0.0684     & 0.0685   &0.0225     &0.0920  \\ 
\enddata
\tablecomments{Statistical properties of the prediction on simulated testing data. From top to bottom we show $R^{2}$; the fraction of outliers and the NMAD for the magnitude $mag$; effective radius $R_{\rm e}$; position angle $PA$; axis ratio $q$ and S\'ersic index $n$. Generally, the log-prediction is better than linear-prediction.}}
\label{tab:stat_1}
\end{deluxetable}

Overall the relative biases (see inset labels) are usually $<10\%$ at all luminosities, while a rather increase of the $\Delta_p$ at $z>1$ is observed, remaining confined within the $20\%$ at $z\sim1.5$ though. All other indicators show reasonably modest values in terms of scatter (NMAD$<0.2$) and outlier fraction ($<0.25$). 

Compared to the result of the GaLNet-2 in Li+22, here we can show the advantages of the space observations in terms of depth and angular size. Indeed, with the SNR$=$40 cut, GaLNet-2 can perform the S\'ersic analysis down to $r\sim 24$, while in Li+22, where we used a more conservative SNR$=$50, we reached $r\sim 22$. In terms of angular size, using space observations
Fig. \ref{fig:pred_1S} shows that also at $z>1$ (see right column)
the accuracy of sub-arcsec effective radii is reasonably good down to $\sim 0.15''$, i.e. $\sim 2\times$ pixel scale (0.074$''$). 
Also the $n$-index show a nice consistency with the ground truth values, although for high-z there is more scatter and outlier fraction, due to the few pixels available for the fit. 

Overall this test shows that GaLNet can push the CSST data to the analysis of the structural parameters for galaxies with a dominant single component to high-$z$ (up to 3). Here we expect to see several millions of galaxies even with the 1-epoch depth, hence providing an unprecedented dataset for understanding the assembly of galaxies in their early phase of their formation (e.g. \citep{2005ApJ...620..564C}; \citep{2018MNRAS.478.3994C})

\begin{figure}
\hspace{-0.7cm}
\includegraphics[width=9.5cm]{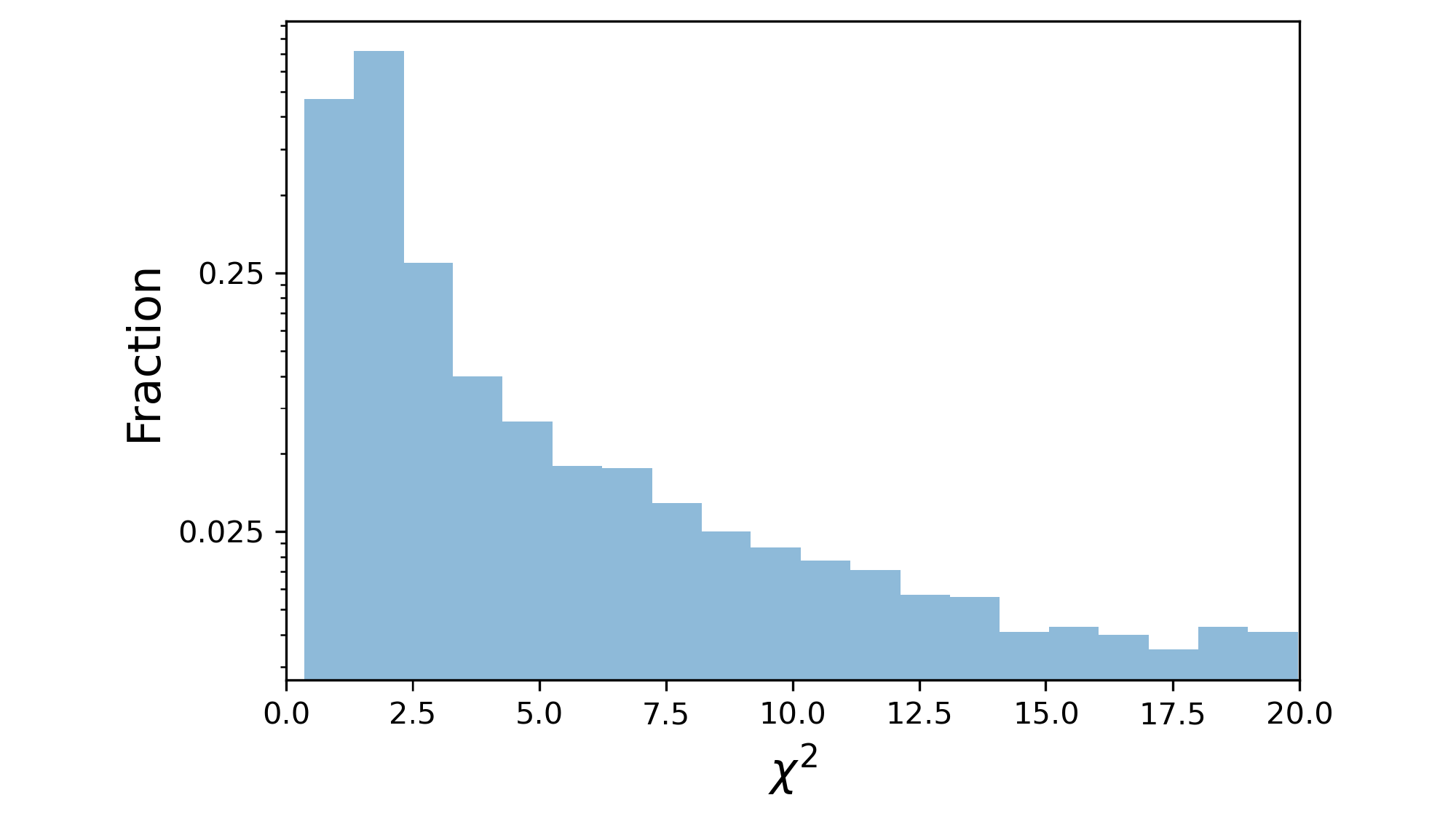}
\caption{Distribution of the $\chi^2$, as defined in Eq. \ref{eq:chi2}, for the GaLNet-2 predictions simulated images. The peak at $\chi^2<2$ contains almost the $\sim$75\%.
\label{fig:chi2_1ser}}
\end{figure}


\section{Summary and Conclusions} \label{sec:concl}
The high-quality and deep data from new generation large sky surveys from space (EUCLID, CSST, Roman)
will give us the opportunity to study the evolution of the morphological mix of galaxies up to early phases of their formation, over unprecedented statistical samples. 
Bulge-Disk structural analysis is a crucial diagnostic to clarify the physics of the galaxy transformation and evolution (e.g. \citep{2005ApJ...620..564C}, \citep{2021ApJ...913..125C}, \citep{2022arXiv221001110F}). However, this is also a fundamental information in cosmological analyses based on weak lensing, as correctly account for the galaxy morphology will prevent strong biases in galaxy shape measurements (e.g. \citep{10.1093/mnras/stu588}). Thus, the multi-component analysis of even billions of galaxies, in different optical and NIR bands, will be a challenging task for upcoming survey programs.

So far, B-D decomposition analyses have been based on traditional analytical tools (e.g., \citep{Gao_2017,2023FrASS..1089443T}), which are computationally demanding. For this reason even in the local universe, these studies are generally limited to small sample of thousands of galaxies (\citep{2020ApJS..247...20Gao2020,2014ApJ...788...11Lang14,2022MNRAS.516..942C}). 


In order to cope with the massive amounts of data generated by the upcoming large-scale observations, deep learning techniques have been proposed to perform parametric galaxy surface brightness analyses. Modeling of single-component, either considering or neglecting the effect of the {\it PSF} (e.g.  \citep{2022ApJ...929..152L} and \citep{2018MNRAS.475..894T}, respectively), have been proved to be a viable alternative to standard techniques, as they provide similar accuracy of standard methods, but with $\sim10^3\times$ faster computational time.


As an evolution of the Galaxy Light profile neural Network (GaLNet) series (see Li+22), we have introduced the first deep learning tool for
the bulge-disk decomposition, dubbed GaLNet-BD. In particular, we have trained the GaLNet-BD to be applied to CSST single-epoch mock simulations, as prototype of optical space observations capable of providing billion galaxy samples. We have considered the measurement of structural parameters, as derived by the 2-S\'ersic profiles, specifically the effective radius, $R_{\rm e}$, the surface brightness at the effective radius, $I_e$, the axis ratio, $q$, the S\'ersic index, $n$, and the position angle, $PA$, for the bulge and the disk respectively. 

To produce realistic galaxies to be used for the training and testing of the new GaLNet-BD sample, we have followed the B/D luminosity, size, axis ratio and redshift from CosmoDC2, and expanded the $n$-index distribution to account for bulges with $n>2$ and disks with $n<2$. We have also tested the case of 1-S\'ersic galaxies, to check the ability of the GaLNets to push the structural parameters analysis to higher redshifts ($\sim1.5$).

We summarize here below the main results of the paper:

\begin{enumerate}
    \item The \GAL-BD can accurately predict the magnitude of the Bulge and the Disk components of galaxies ($R^2\gsim0.87$) with NMAD$\sim0.1$ and typical precision of the order of $\sim20\%$, implying a robust {estimate} of the bulge over total (B/T), although the accuracy of this latter parameter is poorer that the one found for the magnitude of the individual components and we observe the presence of some systematics at low B/T. If considering a ``bright'' sample made of systems with $mag_{\rm disk}$ and $mag_{\rm bulge}$ brighter than $r=22$, then the B/T parameter shows a slightly improved accuracy and no sign of systematics.  
    
    \item High accuracies ($R^2\gsim0.9$) are found for the effective radius, $R_{\rm e}$, the position angle $PA$ of both components. For the S\'ersic index we have found mixed results as the ones of the bulge, $n_{\rm bulge}$, is recovered extremely well ($R^2\gsim0.98$), while the ones of the disk, $n_{\rm disk}$, shows a large scatter and a relative bias at low ($n_{\rm disk}<0.75$) and high end ($n_{\rm disk}>1.50$) of its distribution. The overall accuracy measured by the $R^2$ is $\sim0.1$.  
     \item The relative bias, NMAD and outlied fraction of the main parameters (namely bulge and disk $n$ and $R_{\rm e}$) as a function of the galaxy magnitudes, SNR and redshift (see Fig. \ref{fig:snr_mag_z}) show that \GAL-BD produces stable performances for SNR$>40$. This is possibly a conservative lower SNR limit that can be pushed down to increase the completeness of the structural parameter catalogs. However all the disgnostics start to degrade 
     at z$\gsim1$, especially the disk $n$-index and the bulge $R_{\rm e}$. $z=1$ eventually represents the current upper limit for robust estimates of the B-D parameters, at least with a single epoch CSST observations.
     \item We have finally trained the GaLNet-2 (Li+22), using 1-S\'ersic mock galaxies and the local PSF, to demonstrate the capability of deep space observations, as the ones expected from the CSST, in pushing the galaxy structural parameter analysis toward fainter galaxies and higher redshift. When considering simpler systems and  a smaller number of parameters. One-component model catalogs naturally complement the ones derived from the B-D decomposition, e.g. for normal spheroids. Furthermore, 1-S\'ersic profiles can be robustly used to study the size evolution of galaxies (e.g. \citep{2007MNRAS.382..109T}, \citep{Wel2014ApJ...792L...6V}) of for quantitative morphology (e.g. using the S\'ersic index, see e.g. \citep{10.1093/mnras/sts150}) up to redshifts well beyond $z>0.5$. In this case, we have found that the GaLNet-2 correctly predict the ground truth values with accuracies of the order of $R^2\sim0.9$ for most of the parameters (see Table \ref{tab:stat_1}). Looking at the dependence of the relative bias, outlier fraction and NMAD as a function of magnitude and redshift, we have found that they start to degrade 
     at $z>1$, but they remain reasonably under control up to $z\sim1.5$ or slightly larger, and down to 
     $r\sim 24$.
\end{enumerate}

To conclude, we have provided the first deep learning tool which is capable of efficient and accurate bulge-disk decomposition of galaxies from optical space observations and specifically optimized for upcoming CSST observations. 
In the era of all sky-surveys, the big advantage of Deep Learning tools is the very easy applicability and scalability, combined with an enormous gain in computation time. 
For instance, after a fairly short training time ($\sim$60 minutes on NVIDIA GeForce RTX 3080 Laptop Graphic Processing Unit -- GPU), GaLNet-BD needs only ~140 seconds for single component and up to ~200 seconds for two-components models, to fully analyse $\sim50$k galaxies. This means that with the use of a limited number of commercial GPUs it is possible to analyse 1 billion galaxy samples in one band in a few days. If one combines data from optical and near infrared bands, for instance from CSST and EUCLID, 1 billion galaxies will become doable in one or two months.

The variety of opportunities offered by this capability is enormous. Besides galaxy formation and evolution science, this will allow us to perform a rapid classification of host systems for spectroscopic follow-ups and/or new discoveries (e.g. transient, gravitational lensing etc.), that needs morphology ``and'' full spectral energy distribution (SED) based on multi-band surface photometry, including a quick characterization of the galaxy stellar population (e.g. age, star-fromation etc.). This would be of great benefit for all the community working on these survey facilities. Science-wise, the study of the morphological mix of galaxies from $z\sim3$ over an unprecedented sample of galaxies, down to $r\sim23$ will put stringent constraints upon the evidences that are emerging from the study of even high-redshift galaxies from the James Web Spatial Telescope (e.g. \citep{2020AAS...23520711L}, \citep{2023ApJ...946L..15K})

\section*{Acknowledgements}
\begin{acknowledgements}
NRN acknowledges financial support from the National Science Foundation of China, Research Fund for International Senior Scientists (grant n. 12150710511), from the research grant from China Manned Space project n. CMS-CSST-2021-A01 and A06 and from the Guangdong provincial fund "Understanding structural evolution of galaxies with machine learning” and Guangdong Science Foundation grant (ID: 2022A1515012251). NRN thanks the students of the course ``Reading Astronomy'' at SYSU, for useful suggestions and criticism on the first manuscript draft. RL acknowledges the support of the National Nature Science Foundation of China (No. 12022306) and the science research grants from the China Manned Space Project (CMS-CSST-2021-A01). CT acknowledges funding from INAF Research Grant 2022 LEMON.
LCH was supported by the National Science Foundation of China (11721303, 11991052, 12011540375, 12233001), the National Key R\&D Program of China (2022YFF0503401), and the China Manned Space Project (CMS-CSST-2021-A04, CMS-CSST-2021-A06).
We thank Cheng Li (Tsinghua University) for useful discussions and suggestions.
\end{acknowledgements}

\bibliography{paper_DB}{}
\bibliographystyle{aasjournal}
\appendix
\section{Flat Prior Distribution for the Mock Galaxies}
\label{sec:app}
In this Appendix we want to briefly discuss the impact of the priors, meaning the parameter distribution of the Bulges an Disks in the simulated galaxies, on the GaLNet-DB predictions' accuracy and precision. The reason to test this is twofold: 1) to check whether the under-sampling of the training size in some position of the parameter space can produce an under-performance of the CNN due to poor learning; 2) whether training on a prior distribution much different from the real one can produce stronger biases on the CNN predictions. Possibly we can check if there is a trade-off ebtween these two effects which we can benefit of. 

The first point is motivated by the fact that scatter of the prediction in the $\Delta p$ and NMAD plot, as well as the number of outliers, increase for the poor sampling of some parameters, like the $n_{\rm bulge}$ at high-$z$ discussed in \S\ref{sec:snr_mag_z}.

The second point has been already of our concern in our first analysis (Li+22), where we have concluded that the prior distribution have a minimal impact if close to the real parameter distribution, but we have also seen some deviation emerging if we adopted a flat distribution for the parameters.

Here we use again a flat distribution both providing an unbalanced coverage of the parameter space and an extremely deviating distribution from the actual one. If this produces better results for all parameter, then we can conclude that, for real applications, one can decide to use this latter to realise a balanced sampling of the parameter space.

\begin{figure}[b]
\centering
\includegraphics[width=14.2cm]{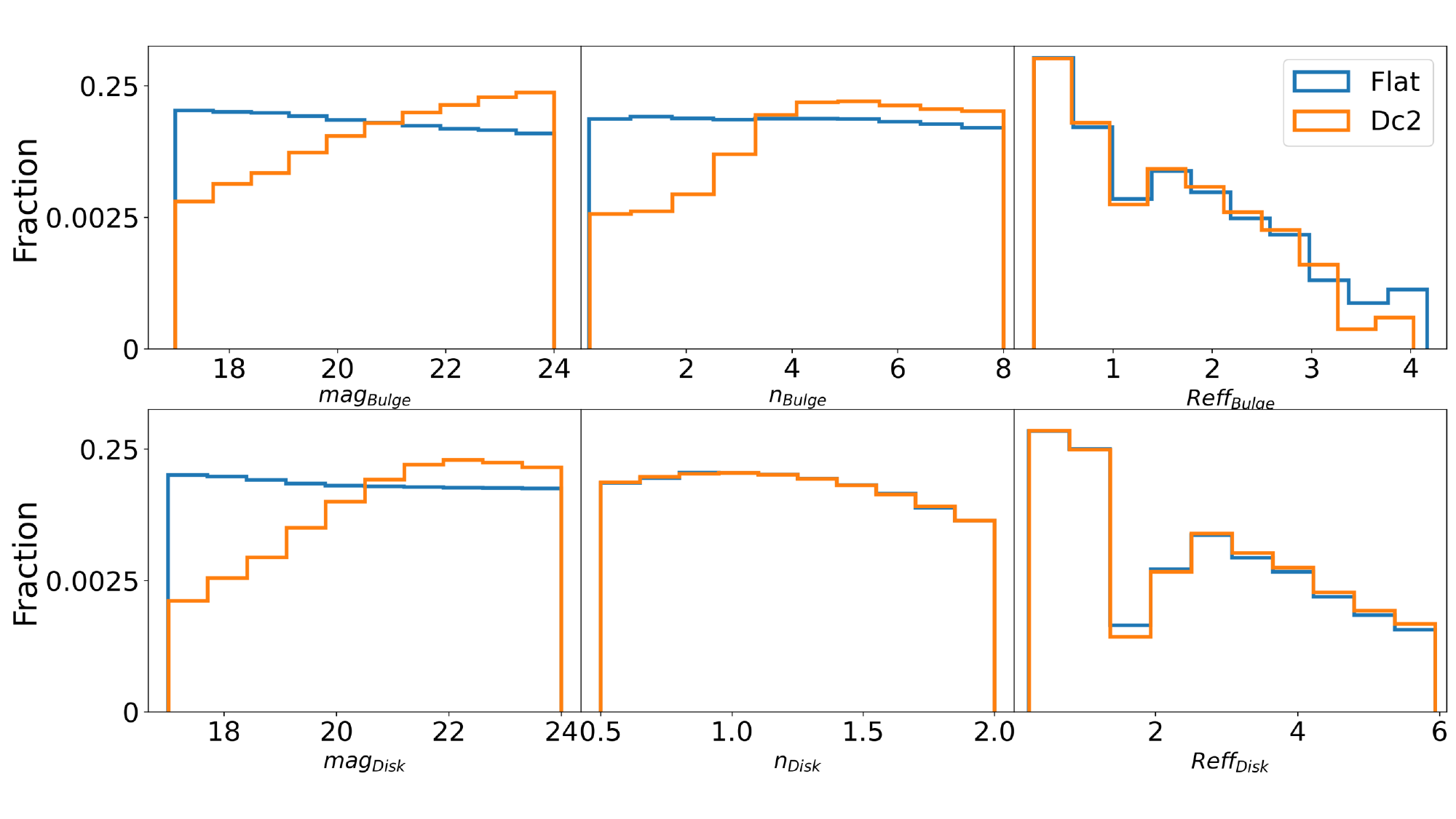}
\caption{Flat "prior" distribution on $mag_{bulge}$ and $mag_{disk}$ to simulate a "flat" dataset.  \label{fig:mag_flat}}
\end{figure}

In particular we have used a flat prior distribution on $mag_{\rm bulge}$ and $mag_{\rm disk}$, by randomly picking galaxies from the CosmoDC2 catalog, with the condition of filling all magnitude bins in a uniform way. Doing this, for the new training sample, we maintain all the physically motivated parameters for each galaxies from CosmoDC2 (except our expanded choice on the $n$-indexes). Hence the final parameter distribution will change as a consequence of the flat distribution of the B and D magnitudes. In Fig. \ref{fig:mag_flat} we show the final distribution of the parameters after having applied the SNR cut, as discussed in \S\ref{sec:SNR_cut}, for the main parameters of the training sample.  
As we have focused this test on magnitude, it is reasonable to modify the priors using the magnitude as a start. However we can see from Fig. \ref{fig:mag_flat} that also the distributions of other parameters become different from the original ones in Fig. \ref{fig:n_distr}. 

\begin{figure}
\centering
\includegraphics[width=18cm]{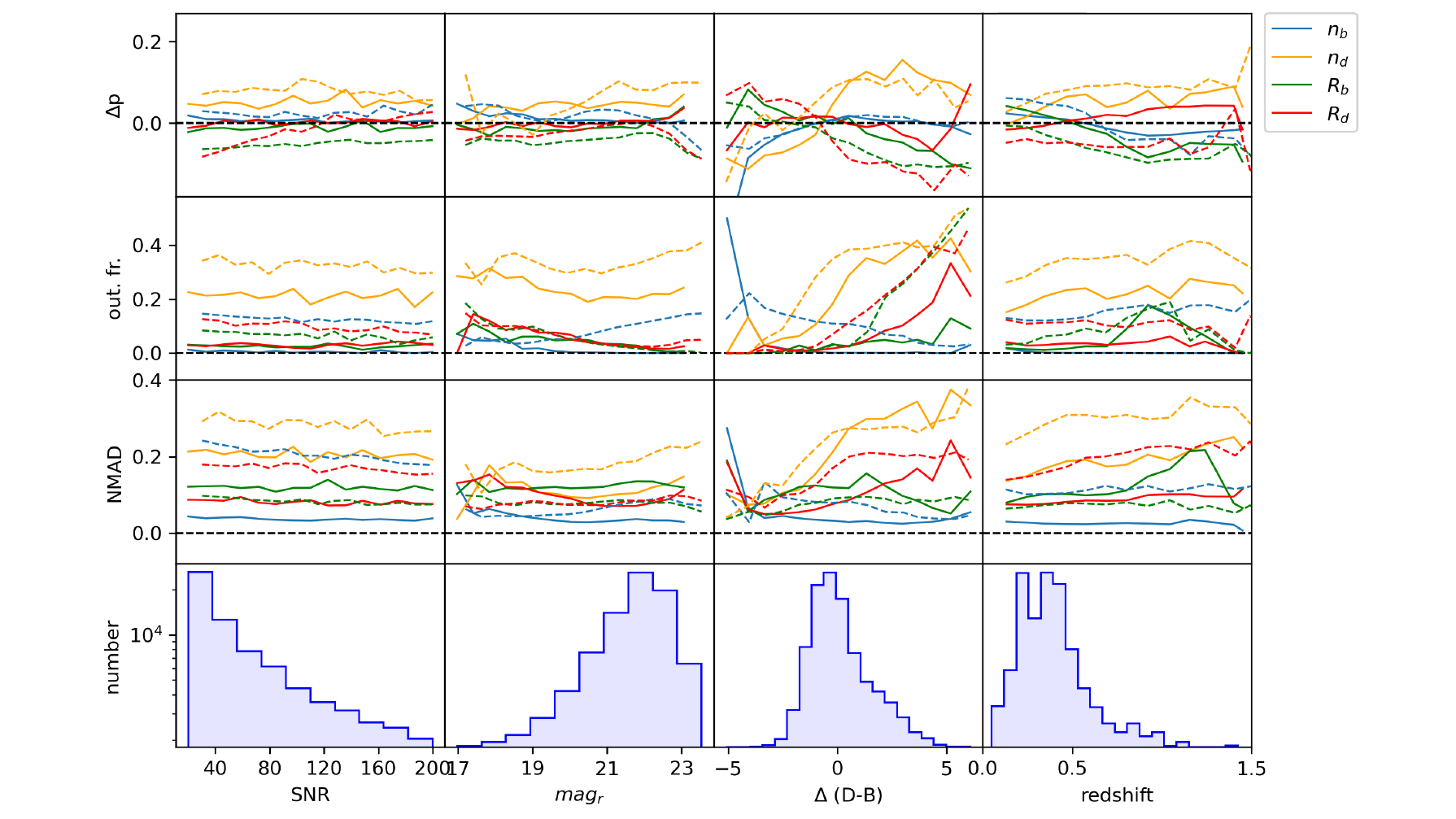}
\caption{Accuracy and precision for 2-component systems with flat priors. Relative bias ($\Delta p$), Outlier fraction (out. fr.), and scatter (NMAD) as functions of SNR, magnitudes, $\Delta$(D-B) and spec-$z$. In the last row we also present the number distribution in the corresponding parameter space. Dashed curves are the results of the ``flat priors'' training while the continuous curves are the from the ``true'' galaxy parameter distribution as in Fig. \ref{fig:snr_mag_z}.  
\label{fig:flat_lines}}
\end{figure}

In Fig. \ref{fig:flat_lines} we can see the result by testing the newly trained \GAL-DB over the same test sample as in \S\ref{sec:perf_linear}, showing the predicted parameters compared to the ground truth. The newly trained GaLNet-DB shows generally a worse accuracy with respect to the one shown in Fig. \ref{fig:snr_mag_z} and reported in the same figure (see for instance the plot against the SNR). We can see some improvement in some parameter, for instance, the $n_{\rm disk}$ as a function of the $\Delta(D-B)$ and toward the brightest magnitudes, or the $n_{\rm bulge}$ toward bright magnitudes and disk dominated systems ($\Delta$(D-B)$<-2.5$). However, the improvement in a limited parameter space volume is mirrored by a general degradation of all other parameters, demonstrating that the deviation from a ``realistic'' parameter distribution of the prior does not produce any advantage in the \GAL-DB predictions.


\end{document}